\documentclass[aps,twocolumn,floats,prd,nofootinbib,10pt,longbibliography,superscriptaddress]{revtex4-1}

\usepackage{comment}
\usepackage[dvips]{graphicx}
\usepackage{graphicx,amsmath,amsfonts,amssymb,slashed,float}
\usepackage[breaklinks=true,colorlinks=true,linkcolor=blue,urlcolor=blue,citecolor=blue]{hyperref}
\usepackage[normalem]{ulem}
\usepackage{bbold,wasysym}
\usepackage{graphicx}
\usepackage{array,multirow}
\usepackage[utf8]{inputenc}
\usepackage[usenames,dvipsnames]{xcolor} 
\usepackage{soul}
\usepackage[pscoord]{eso-pic}
\usepackage{makecell}
\usepackage{placeins}

\setstcolor{Blue}

\def\aap{A\&A}

\newcommand{\LCDM}{$\Lambda$CDM}
\newcommand{\ACT}{ACT DR4}
\newcommand{\SHOES}{S$H_0$ES}

\newcommand{\placetextbox}[3]{
	\setbox0=\hbox{#3}
	\AddToShipoutPictureFG*{
		\put(\LenToUnit{#1\paperwidth},\LenToUnit{#2\paperheight}){\vtop{{\null}\makebox[0pt][c]{#3}}}
	}
}

\begin{document}

\placetextbox{0.85}{0.97}{\small ULB-TH/22-03, LUPM:22-003}

\title{Hints of Early Dark Energy in \textit{Planck}, SPT, and ACT data: \\ new physics or systematics?}
\author{Tristan L.~Smith}
\affiliation{Department of Physics and Astronomy, Swarthmore College, Swarthmore, PA 19081, USA}
\author{Matteo Lucca}
\affiliation{Service de Physique Th\'{e}orique, Universit\'{e} Libre de Bruxelles, C.P. 225, B-1050 Brussels, Belgium}
\author{Vivian Poulin}
\affiliation{Laboratoire Univers \& Particules de Montpellier (LUPM), CNRS \& Universit\'{e} de Montpellier (UMR-5299), Place Eug\`{e}ne Bataillon, F-34095 Montpellier Cedex 05, France}
\author{Guillermo F. Abellan}
\affiliation{Laboratoire Univers \& Particules de Montpellier (LUPM), CNRS \& Universit\'{e} de Montpellier (UMR-5299), Place Eug\`{e}ne Bataillon, F-34095 Montpellier Cedex 05, France}
\author{Lennart Balkenhol}
\affiliation{School of Physics, University of Melbourne, Parkville, VIC 3010, Australia}
\author{Karim Benabed}
\affiliation{Sorbonne Universit\'e, CNRS, UMR 7095, Institut d'Astrophysique de Paris, 98 bis bd Arago, 75014 Paris, France}
\author{Silvia Galli}
\affiliation{Sorbonne Universit\'e, CNRS, UMR 7095, Institut d'Astrophysique de Paris, 98 bis bd Arago, 75014 Paris, France}
\author{Riccardo Murgia}
\affiliation{Laboratoire Univers \& Particules de Montpellier (LUPM), CNRS \& Universit\'{e} de Montpellier (UMR-5299), Place Eug\`{e}ne Bataillon, F-34095 Montpellier Cedex 05, France}
\affiliation{Gran Sasso Science Institute (GSSI), I-67100 L’Aquila (AQ), Italy}
\affiliation{INFN - Laboratori Nazionali del Gran Sasso (LNGS), I-67100 L’Aquila (AQ), Italy}

\begin{abstract}
We investigate constraints on early dark energy (EDE) using \ACT{}, SPT-3G 2018, {\it Planck} polarization, and restricted {\it Planck} temperature data (at $\ell<650$), finding a $3.3\sigma$ preference ($\Delta\chi^2=-16.2$ for three additional degrees of freedom) for EDE over $\Lambda$CDM. The EDE contributes a maximum fractional energy density of $f_{\rm EDE}(z_c) = 0.163^{+0.047}_{-0.04}$ at a redshift $z_c = 3357\pm 200$ and leads to a CMB inferred value of the Hubble constant $H_0 = 74.2^{+1.9}_{-2.1}$ km/s/Mpc. We find that  {\it Planck} and \ACT{} data provide the majority of the improvement in $\chi^2$, and that the inclusion of SPT-3G pulls the posterior of $f_{\rm EDE}(z_c)$ away from $\Lambda$CDM. This is the first time that a moderate preference for EDE has been reported for these combined CMB data sets including Planck polarization. We find that including measurements of supernovae luminosity distances and the baryon acoustic oscillation standard ruler only minimally affects the preference ($3.0\sigma$), while measurements that probe the clustering of matter at late times -- the lensing potential power spectrum from {\it Planck} and $f \sigma_8$ from BOSS -- decrease the significance of the preference to 2.6$\sigma$. Conversely, adding a prior on the $H_0$ value as reported by the \SHOES{} collaboration increases the preference to the $4-5\sigma$ level. In the absence of this prior, the inclusion of \textit{Planck} TT data at $\ell>1300$ reduces the preference from $3.0\sigma$ to $2.3\sigma$ and the constraint on $f_{\rm EDE}(z_c)$ becomes compatible with \LCDM{} at $1\sigma$. We explore whether systematic errors in the \textit{Planck} polarization data may affect our conclusions and find that changing the TE polarization efficiencies significantly reduces the \textit{Planck} preference for EDE. More work will be necessary to establish whether these hints for EDE within CMB data alone are the sole results of systematic errors or an opening to new physics.
\end{abstract}

\maketitle

\section{Introduction}\label{sec: intro}
Over the past several years, the standard cosmological model, \LCDM, has come under increased scrutiny as measurements of the late-time expansion history of the Universe~\cite{Scolnic:2017caz}, the cosmic microwave background (CMB)~\cite{Planck:2018vyg}, and large-scale structure (LSS) -- such as the clustering of galaxies~\cite{Alam:2016hwk,Abbott:2017wau,Hildebrandt:2018yau,eBOSS_cosmo} -- have improved. Some of these observations have hinted at possible tensions within \LCDM{}, related to the Hubble constant ${H_0 = 100 h}$~km/s/Mpc~\cite{Verde:2019ivm} and the parameter combination $S_8 \equiv \sigma_8(\Omega_{\rm m}/0.3)^{0.5}$~\cite{Joudaki:2019pmv} (where $\Omega_{\rm m}$ is the total matter relic density parameter and $\sigma_8$ is the root mean square of the linear matter perturbations within 8 Mpc/$h$ today), reaching the $4-5\sigma$ \cite{Riess:2020fzl,Soltis:2020gpl,Pesce:2020xfe, Blakeslee:2021rqi, Riess:2021jrx} and $2-3\sigma$ level \cite{Joudaki:2019pmv,Heymans:2020gsg,DES:2021wwk}, respectively. While both of these discrepancies may be the result of systematic uncertainties, and not all measurements lead to the same level of tension \cite{Freedman:2019jwv,Freedman:2020dne} (see also Refs. \cite{Freedman:2021ahq,Anand:2021sum}), numerous models have been suggested as a potential resolution (see e.g. Refs.~\cite{DiValentino:2021izs,Schoneberg:2021qvd} for recent reviews), though none is able to resolve both tensions simultaneously \cite{Jedamzik:2020zmd,Schoneberg:2021qvd}. 

In this work we focus on a scalar field model of `early dark energy' (EDE), originally proposed to resolve the `Hubble tension' (see e.g. Refs.~\cite{Karwal:2016vyq,Poulin:2018dzj, Poulin:2018cxd,  Smith:2019ihp}). The EDE scenario assumes the presence of an ultra-light scalar field $\phi$ slow-rolling down an axion-like potential of the form $V(\phi)\propto [1-\cos(\phi/f)]^n$, where $f$ is the decay constant of the field. Due to Hubble friction the field is initially fixed at some value, $\theta_i = \phi_i/f$, and becomes dynamical when the Hubble parameter drops below the field's mass, which happens at a critical redshift $z_c$. Once that occurs, the field starts to evolve, eventually oscillates around the minimum of its potential, and its energy density dilutes at a rate faster than matter (for the potential we use here, with $n=3$, $\rho_{\rm EDE} \propto (1+z)^{4.5}$). The energy density of the scalar field around $z_c$ reduces the sound horizon at recombination leading to an increase in the inferred value of $H_0$ from CMB measurements (see e.g. Ref.~\cite{Knox:2019rjx}).

Up until recently, evidence for EDE came only from analyses which included a prior on the value of $H_0$ from the Supernova $H_0$ for the Equation of State (\SHOES) collaboration\footnote{The \SHOES{} prior is actually a constraint on the absolute calibration of the SNe data. However, since the EDE is dynamical at pre-recombination times, this distinction is unimportant~\cite{Schoneberg:2021qvd}.} \cite{Poulin:2018cxd, Smith:2019ihp, Chudaykin:2020acu,Chudaykin:2020igl,Murgia:2020ryi}.~Using this prior on $H_0$ and the full \textit{Planck} power spectra, within the EDE model one obtains a non-zero fraction of the total energy density in EDE at the critical redshift, ${f_{\rm EDE}(z_c) = 0.108^{+0.035}_{- 0.028}}$\,, with a corresponding Hubble parameter ${H_0 = 71.5 \pm 1.2}$ km/s/Mpc~\cite{Murgia:2020ryi} (adding supernovae (SNe) and baryon acoustic oscillation `standard ruler' (BAO) data leads to insignificant shifts). Without the \SHOES{} prior, one has instead an upper bound of the form ${f_{\rm EDE}(z_c) < 0.088}$ at 95\% confidence level (CL) and  ${H_0 = 68.29^{+0.75}_{-1.3}}$~km/s/Mpc~\cite{Hill:2020osr,Murgia:2020ryi}.\footnote{Given the weak evidence for EDE, the marginalized constraints are strongly dependent on the choice of priors for the EDE parameters, making these constraints hard to interpret \cite{Murgia:2020ryi,Smith:2020rxx,Herold:2021ksg}.}

Recent analyses of EDE using data from the Atacama Cosmology Telescope's fourth data release (\ACT{})~\cite{ACT:2020frw} alone have shown a slight ($\sim 2.2\sigma$) preference for the presence of an EDE component with a fraction $f_{\rm EDE}(z_c)\sim 0.15$ and $H_0 \sim 74$ km/Mpc/s \cite{Hill:2021yec,Poulin:2021bjr}. Interestingly, the inclusion of large-scale CMB temperature measurements by the Wilkinson Microwave Anisotropy Probe (WMAP) \cite{WMAP:2012fli} or the \textit{Planck} satellite~\cite{Planck:2018vyg} restricted to the WMAP multipole range increases the preference to $\sim3\sigma$. A similar analysis using the third generation South Pole Telescope 2018 (SPT-3G) data \cite{SPT-3G:2021eoc} was presented in Ref.~\cite{LaPosta:2021pgm} (see also Refs.~\cite{Chudaykin:2020acu,Chudaykin:2020igl} for previous studies using SPTpol). There is no evidence for EDE over \LCDM{} using SPT-3G alone or when combined with the \textit{Planck} temperature power spectrum restricted to $\ell<650$, giving the marginalized constraint $f_{\rm EDE}(z_c)<0.2$ at 95\% CL in the latter case. Combining \ACT{} and/or SPT-3G with the full \textit{Planck} CMB power spectra returns an upper limit on $f_{\rm EDE}(z_c)$, albeit less restrictive than for \textit{Planck} alone.

In Refs.~\cite{Hill:2021yec,Poulin:2021bjr} it was argued that the \ACT{} preference for EDE is mainly driven by a feature in the \ACT\ EE power spectrum around $ \ell \sim 500$  when \ACT{} is considered alone, with an additional broadly-distributed contribution from the TE spectrum when in combination with restricted \textit{Planck} TT data ($\ell <650$ or $\ell <1060$). Ref.~\cite{Poulin:2021bjr} also considered the role of \textit{Planck} polarization data, finding that the evidence for a non-zero $f_{\rm EDE}(z_c)$ and an increased $H_0$ persists, as long as the {\em Planck} TT spectrum is restricted to $\ell < 1060$. 

Building on these previous studies, the work presented here explores in more detail how the evidence for EDE using data from \ACT{}, SPT-3G or both data sets is impacted by the inclusion of the more precise intermediate-scale ($\mathcal{O}(\ell) = 100$) polarization measurements by \textit{Planck}. We test the robustness of the results to changes in the \textit{Planck} TE polarization efficiency and the dust contamination amplitudes in \textit{Planck} EE. We also further investigate the role of \textit{Planck} high-$\ell$ TT data as well as that of several non-CMB probes.

This paper is organized as follows. In Section \ref{sec:ana} we briefly summarize the numerical setup and cosmological data sets used in our analysis. In Section~\ref{sec:results} we present our results, focusing on the role of \textit{Planck} polarization and temperature data as well as that of possible systematic uncertainties. We conclude in Section~\ref{sec:concl} with a summary and final remarks. The Appendices contain additional figures and tables. 

\section{Analysis method and data sets}\label{sec:ana}
For the numerical evaluation of the cosmological constraints on the models considered within this work (\LCDM{} and EDE) and their statistical comparison we perform a series of Markov-chain Monte Carlo (MCMC) runs using the public code {\sf MontePython-v3}\footnote{\url{https://github.com/brinckmann/montepython_public}} \citep{Audren:2012wb,Brinckmann:2018cvx}, interfaced with our modified version\footnote{\url{https://github.com/PoulinV/AxiCLASS}} of {\sf CLASS}\footnote{\url{https://lesgourg.github.io/class_public/class.html}} \cite{Lesgourgues:2011re,Blas:2011rf}. We make use of a Metropolis-Hasting algorithm assuming uninformative flat priors on $\{\omega_b,\omega_{\rm cdm},H_0,A_s,n_s,\tau_{\rm reio}\}$\footnote{Here $\omega_b$ and $\omega_{\rm cdm}$ are the physical baryon and cold DM energy densities, respectively, $A_s$ is the amplitude of the scalar perturbations, $n_s$ is the scalar spectral index, and $\tau_{\rm reio}$ is the reionization optical depth.}, while when considering the EDE model we also vary $\{\log_{10}(z_c),f_{\rm EDE}(z_c) ,\theta_i\}$ with priors\footnote{We focus on the range of $z_c$ for which EDE mostly affects the sound horizon, and therefore $H_0$. Broadening the $z_c$ range can affect the constraints on $f_{\rm EDE}(z_c)$ from SPT-3G alone or in combination with \textit{Planck} TT650 \cite{LaPosta:2021pgm}.} of the form $\{3 \leq \log_{10}(z_c)\leq 4, 0.001 \leq f_{\rm EDE}(z_c) \leq 0.5 ,0.01 \leq \theta_i \leq 3.1\}$. We also include all nuisance parameters associated with each data set as given by the official collaborations and treat the corresponding sets of nuisance parameters independently.\footnote{In principle, one could use a common foreground model, which would reduce the overall number of free parameters and possibly reduce the uncertainties on the cosmological parameters. However, the publicly available likelihoods do not easily allow this and therefore many (if not all) joint CMB analyses that have appeared in the literature employ a separate foreground modeling (see, e.g., Refs.~\cite{ACT:2020gnv,SPT-3G:2021eoc}). Furthermore, our analysis shows that the posterior distributions for the foreground parameters are identical in the $\Lambda$CDM and EDE cosmologies and that they are uncorrelated with the EDE parameters. Because of this, we do not expect a joint foreground model to have a significant impact on our results.}  As described in Ref.~\cite{Smith:2019ihp}, we use a shooting method to map the set of phenomenological parameters $\{\log_{10}(z_c), f_{\rm EDE}(z_c)\}$ to the theory parameters $\{m,f\}$. We adopt the {\em Planck} collaboration convention in modeling free-streaming neutrinos as two massless species and one massive with $m_\nu=0.06$ eV \cite{Ade:2018sbj}, and use {\sf Halofit} to estimate the non-linear matter clustering \cite{Smith:2002dz}. We consider chains to be converged using the Gelman-Rubin \citep{Gelman:1992zz} criterion $|R -1|\lesssim0.05$.\footnote{This condition is chosen because of the non-Gaussian (and sometimes multi-modal) shape of the posteriors of the EDE parameters. For all \LCDM{} runs we have $|R -1|<0.01$.} To post-process the chains and produce our figures we use {\sf GetDist} \cite{Lewis:2019xzd}.

\begin{figure}
    \centering
    \includegraphics[width=0.85\columnwidth]{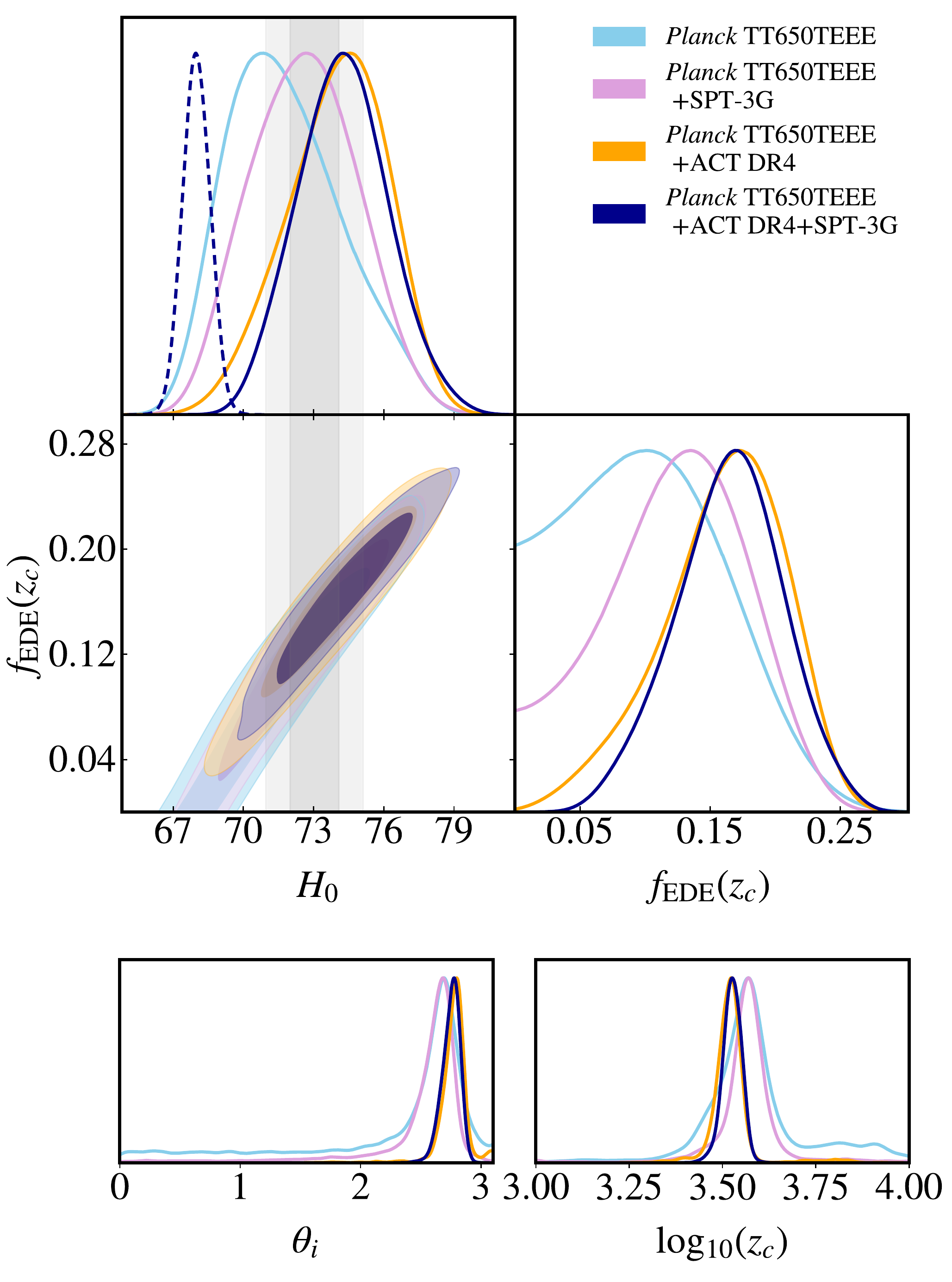}
    \caption{1D and 2D posterior distributions (68\% and 95\% CL) of $H_0, f_{\rm EDE}(z_c), \theta_i,$ and $\log_{10}(z_c)$ for different data set combinations. The vertical gray band shows $H_0=73.04 \pm 1.04$ km/s/Mpc, as reported by the \SHOES{} collaboration \cite{Riess:2021jrx}. The dashed curve shows the posterior distribution for $H_0$ within \LCDM{} with \textit{Planck} TT650TEEE+\ACT+SPT-3G. When all data sets are combined, EDE is preferred at the $\sim 3\,\sigma$ level and leads to a higher $H_0$ value, in good agreement with the \SHOES{} result.}
    \label{fig: MCMC_res}
\end{figure}

We make use of the various \textit{Planck} 2018 \cite{Planck:2018vyg} and \ACT\  \cite{ACT:2020frw} likelihoods distributed together with the public {\sf MontePython} code, while the SPT-3G polarization likelihood \cite{SPT-3G:2021eoc} has been adapted from the official {\sf clik} format\footnote{\url{https://pole.uchicago.edu/public/data/dutcher21} (v3.0)}. In addition to the full \textit{Planck} polarization power spectra (refered to as TEEE), we compare the use of the \textit{Planck} TT power spectrum with a multipole range restricted to $\ell<650$ (TT650), or the full multipole range (TT). The choice of \textit{Planck} TT650 is motivated by the fact that the \textit{Planck} and WMAP data are in excellent agreement in this multipole range \cite{huang2018}. In all the runs of this paper, we include the \textit{Planck} low multipole ($\ell<30$) EE likelihood to constrain the optical depth to reionization, as well as the low-$\ell$ TT likelihood \cite{Planck:2018vyg}. For any data combination that includes \textit{Planck} TT650 we did not restrict \ACT{} TT. In analyses that include \textit{Planck} TT at higher multipoles, we removed any overlap with \ACT{} TT up until $\ell = 1800$ to avoid introducing correlations between the two data sets \cite{Aiola:2020azj}. 

Finally, we briefly explore joint constraints from the primary CMB anisotropy data in combination with CMB lensing potential measurements from \textit{Planck} \cite{Planck:2018vyg}, BAO data gathered from 6dFGS at $z = 0.106$ \cite{Beutler:2011hx}, SDSS DR7 at $z = 0.15$ \cite{Ross:2014qpa} and BOSS DR12 at ${z = 0.38, 0.51, 0.61}$ \cite{Alam:2016hwk} (both with and without information on redshift space distortions (RSD) $f\sigma_8$), data from the Pantheon catalog of uncalibrated luminosity distance of SNe in the range ${0.01<z<2.3}$~\cite{Scolnic:2017caz} as well as the late-time measurement of the $H_0$ value reported by the \SHOES{} collaboration, $H_0=73.04\pm1.04$ km/s/Mpc \cite{Riess:2021jrx} (which we account for as a Gaussian prior on $H_0$).

\begin{table}
\def\arraystretch{1.2}
    \scalebox{0.85}{
    \begin{tabular}{|l|c|c|} 
        \hline
        Model & $\Lambda$CDM & EDE \\
        \hline
        \hline
        $f_{\rm EDE}(z_c)$ & $-$ & $0.163(0.179)_{-0.04}^{+0.047}$ \\
        $\log_{10}(z_c)$&$-$ & $3.526(3.528)_{-0.024}^{+0.028}$\\
        $\theta_i$ &  $-$ & $2.784(2.806)_{-0.093}^{+0.098}$ \\
        \hline
        $m$ (eV) &  $-$ & $(4.38 \pm 0.49) \times 10^{-28}$ \\
        $f$ (Mpl) &  $-$ & $0.213 \pm 0.035$ \\
        \hline
        $H_0$ [km/s/Mpc]&$68.02(67.81)_{-0.6}^{+0.64}$  & $74.2(74.83)_{-2.1}^{+1.9}$ \\
        $100~\omega_b$ & $2.253(2.249)_{-0.013}^{+0.014}$  & $2.279(2.278)_{-0.02}^{+0.018}$ \\
        $\omega_{\rm cdm}$ &$0.1186(0.1191)_{-0.0015}^{+0.0014}$ & $0.1356(0.1372)_{-0.0059}^{+0.0053}$\\
        $10^{9}A_s$ & $2.088(2.092)_{-0.033}^{+0.035}$ & $2.145(2.146)_{-0.04}^{+0.041}$ \\
        $n_s$&  $0.9764(0.9747)_{-0.0047}^{+0.0046}$ &  $1.001(1.003)_{-0.0096}^{+0.0091}$ \\
        $\tau_{\rm reio}$& $0.0510(0.0510)_{-0.0078}^{+0.0087}$ & $0.0527(0.052)_{-0.0084}^{+0.0086}$ \\
        \hline
        $S_8$ &$0.817(0.821)\pm0.017$ & $0.829(0.829)_{-0.019}^{+0.017}$\\
        $\Omega_m$ &$0.307(0.309)_{-0.009}^{+0.008}$ & $0.289(0.287)\pm0.009$\\
        Age [Gyrs] & $13.77(13.78)\pm0.023$&$12.84(12.75)\pm0.27$ \\
        \hline
        $\Delta \chi^2_{\rm min}$ (EDE$-\Lambda$CDM) & $-$ &-16.2 \\
        Preference over $\Lambda$CDM &$-$ & 99.9\% ($3.3\sigma$) \\
        \hline
    \end{tabular} }
    \caption{The mean (best-fit) $\pm 1\sigma$ errors of the cosmological parameters reconstructed in the $\Lambda$CDM and EDE models from the analysis of the \ACT{}+SPT-3G+\textit{Planck} TT650TEEE data set combination.}
    \label{tab:full}
\end{table}

\begin{figure*}
    \centering
    \includegraphics[width=1.45\columnwidth]{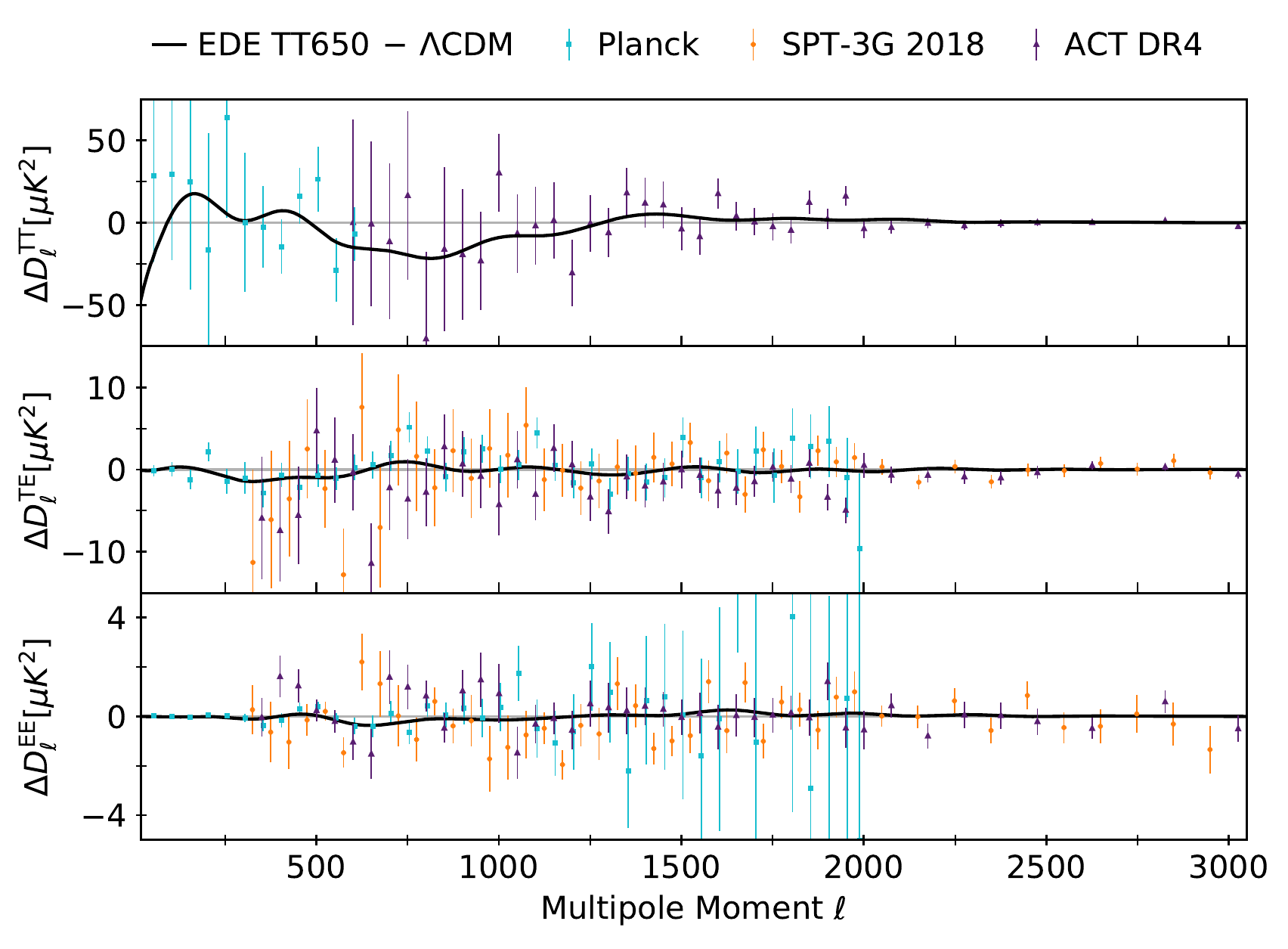}
    \caption{The difference between the EDE and \LCDM{} best-fit models to the data combination \ACT{}+SPT-3G+\textit{Planck} TT650TEEE (solid black) and the residuals of the data points computed with respect to the \LCDM{} best fit of the same data set combination (coloured data points). Although the EE power spectrum measurements around $\ell \sim 500$ of SPT-3G and \textit{Planck} do not follow the same fluctuations as the ACT DR4 data, we find a $3.3\,\sigma$ preference for EDE over \LCDM{} when fitting \ACT{}+SPT-3G+\textit{Planck} TT650TEEE jointly.}
    \label{fig: residuals}
\end{figure*}

\section{Results} \label{sec:results}
The resulting posterior distributions of the parameters most relevant for our discussion are shown in Fig.~\ref{fig: MCMC_res} for a variety of CMB data set combinations. The mean, best-fit, and 1$\sigma$ errors for the full CMB data set combination for both \LCDM{} and EDE cosmologies are shown in Table~\ref{tab:full}. A complete list of CMB constraints can be found in Table \ref{tab:CMB} provided in Appendix~\ref{app: tables}.

We find that the combination of \textit{Planck} TT650TEEE+ \ACT{}+SPT-3G leads to a $3.3\sigma$ preference\footnote{We compute the preference assuming that the $\Delta\chi^2$ follows a $\chi^2$ distribution with three degrees of freedom. Because the parameters $\{z_c,\theta_i\}$ are not defined once $f_{\rm EDE}=0$, this test-statistics does not fully encapsulate the true significance, as required by Wilks' theorem \cite{Wilks:1938dza}. Still, we note that it gives results more conservative than {\em local significance} tests, which would consist in computing the preference at fixed $\{z_c,\theta_i\}$, and therefore with a single degree of freedom.  We keep a more detailed analysis estimating the true significance for future work, for instance following Refs.~\cite{Gross:2010qma,Ranucci:2012ed,Bayer:2020pva} or dedicated mock data analyses.} for EDE over $\Lambda$CDM (${\Delta\chi^2 \equiv \chi^2({\rm EDE})-\chi^2(\Lambda{\rm CDM})= -16.2}$), with $f_{\rm EDE}(z_c)=0.163_{-0.04}^{+0.047}$ and $H_0=74.2^{+1.9}_{-2.1}$ km/s/Mpc (see Table~\ref{tab:full}). Although the $\chi^2$ preference is mainly driven by an improvement of the fit to {\it Planck} TT650TEEE and \ACT{} (the detailed breakdowns of the $\chi^2$ values are given in Appendix~\ref{app: chi2}), the addition of SPT-3G pulls the $f_{\rm EDE}(z_c)$ posterior up relative to {\it Planck} TT650TEEE alone (see Table \ref{tab:CMB}). It is remarkable that with the inclusion of the \textit{Planck} polarization power spectra, hints for the EDE cosmology are present when combined with \textit{Planck} TT650 (2.2$\sigma$), \textit{Planck} TT650+\ACT{} (3.3$\sigma$), \textit{Planck} TT650+SPT-3G (2.4$\sigma$), and when all three data sets are combined (3.3$\sigma$). Moreover, the resulting posterior distributions for $f_{\rm EDE}(z_c)$ visually agree with one another as shown in Figs.~\ref{fig: MCMC_res} and~\ref{fig: MCMC_full}, though quantifying this consistency is complicated by the partly shared data.

\subsection{Impact of \textit{Planck} TEEE data}
In the context of the EDE scenario, it was argued in Refs.~\cite{Hill:2021yec,Poulin:2021bjr} that the preference for a non-zero $f_{\rm EDE}(z_c)$ using \ACT\ data alone or with additional \textit{Planck} low-$\ell$ temperature data is driven, in part, by features in the \ACT{} EE power spectrum around $\ell \sim 500$. The lack of such a feature in the SPT-3G data might explain why in combination with \ACT\, these data do not show evidence for a non-zero $f_{\rm EDE}(z_c)$ \cite{LaPosta:2021pgm}. The effect of adding the \textit{Planck} polarization power spectra is most apparent at the intermediate TE and EE multipoles, since it is at these scales that the \textit{Planck} measurements are more constraining than those of \ACT{} and SPT-3G. 

We show the difference of the TT, TE and EE power spectra between the EDE and \LCDM{} best-fit models extracted from the data set combination \ACT{}+SPT-3G+\textit{Planck} TT650TEEE in Fig.~\ref{fig: residuals}, while in Fig.~\ref{fig: residuals_full} of Appendix \ref{app:nopol} we focus on \ACT{} and SPT-3G data with and without {\em Planck} polarization data. The figures show that \textit{Planck} TEEE data drive tight constraints on the spectra at low multipoles, with a small deviation away from $\Lambda$CDM in TE between $\ell\sim200-800$ and in EE between $\ell\sim500-800$ that is coherent with the behavior of the data. Remarkably, after the inclusion of \textit{Planck} polarization data, the best-fit models for \ACT{} and SPT-3G come into better agreement. Additionally, due to the presence of EDE\footnote{For discussions about the impact of EDE on the CMB power spectra and the correlation with other cosmological parameters see Refs.~\cite{Poulin:2018cxd,Knox:2019rjx,Hill:2020osr,Vagnozzi:2021gjh}.} the TT spectrum exhibits a lower power than $\Lambda$CDM around $\ell\sim 500-1300$, which follows a trend clearly visible in \ACT{} data. In fact, in this combined analysis of ACT DR4 with {\em Planck} TT650TEEE and SPT-3G, the preference for EDE within \ACT{} data is driven almost equally by temperature ($\Delta\chi^2$ (\ACT{} TT)$=-3.3$) and polarization ($\Delta\chi^2$ (\ACT{} TEEE)$=-4.7$) data.

At the parameter level, the main impact of including \textit{Planck} TEEE in combination with \textit{Planck} TT650+\ACT{}+SPT-3G is on the value of $\omega_b$, $z_c$, and $\theta_i$ (for comparison, see Appendix \ref{app:nopol} for analyses without \textit{Planck} polarization data). For instance, \ACT{}+\textit{Planck} TT650 gives $10^{2}\omega_b =  2.154^{+0.04}_{-0.046}$, $\log_{10}(z_c)=3.21^{+0.11}_{-0.01}$ and no constraints on $\theta_i$ (see Table \ref{tab:TTonly} and Fig.~\ref{fig:PlanckTT650}). The inclusion of \textit{Planck} polarization shifts the baryon density to $10^{2}\omega_{b } = 2.273 ^{+0.02}_{-0.023}$, tightly constrains ${\theta_i=2.784^{+0.098}_{-0.093}}$, and leads to a value of the critical redshift $z_c$ in good agreement with that of earlier findings \cite{Poulin:2018cxd,Smith:2019ihp,Murgia:2020ryi}, namely $\log_{10}(z_c)=3.529^{+0.03}_{-0.049}$, i.e. a field that becomes dynamical around the time of matter-radiation equality ($\log_{10}(z_{\rm eq}) = 3.580^{+0.022}_{-0.016}$).

Although there is an overall improvement in the $\chi^2$ when using EDE for all of the CMB data, the inclusion of \textit{Planck} polarization leads to a degradation of the fit to \ACT\: when compared to the EDE analysis with \ACT{}+\textit{Planck} TT650, ${\Delta \chi^2_{\rm ACT} =+11.8}$ (see Tables \ref{tab:chi2_ede} and \ref{tab:chi2_nopol}).\footnote{Even with this increase in the \ACT{} $\chi^2$, the overall goodness-of-fit as quantified by the probability-to-exceed goes from 0.17 to 0.07. Thus, in terms of the overall goodness-of-fit, both models are acceptable.} It is however remarkable that, regardless of the data combination, the improvement over $\Lambda$CDM is similar ($\Delta \chi^2\sim-8$). In the combined fit, we note that the $\chi^2$ of SPT-3G and \textit{Planck} TT650TEEE are also mildly degraded (both in the EDE and $\Lambda$CDM model), and exploring whether these shifts are compatible with pure statistical effects is left for future work (see Ref.~\cite{Handley:2019wlz} for a related discussion).

\subsection{Systematic uncertainty in \textit{Planck} TEEE data}
\label{sec:systs}
As explained in Ref.~\cite{Planck:2019nip} (see also Sec.~2.2.1 of Ref. \citep{Planck:2018vyg} and Ref. \citep{galli2021}), two different approaches for the modeling of the \textit{Planck} TE polarization efficiency (PE) calibration are possible\footnote{Polarization efficiencies are calibration factors multiplying polarization spectra. In principle, the polarization efficiencies found by fitting the TE spectra should be consistent with those obtained from EE. However, in \textit{Planck}, small differences (at the level of $2\sigma$) are found between the two estimates at 143 GHz. There are two possible choices: the `map-based' approach, which adopts the estimates from EE (which are about a factor of 2 more precise than TE) for both the TE and EE spectra; or the `spectrum-based' approach, which applies independent estimates from TE and EE. The baseline \textit{Planck} likelihood uses a `map-based' approach, but allows one to test the `spectrum-based' approach as well (see also Ref. \citep{camspec}), as we do in this paper.}. In principle, these techniques should give equivalent results for the TE PE parameters, but in practice estimates in \textit{Planck} are slightly discrepant, at the level of  $\sim 2\sigma$ (see Eqs.~(45) -- used as baseline -- and (47) of Ref. \cite{Planck:2019nip}). Although these differences have a negligible impact on the parameter estimation within \LCDM{}, it has been noted that constraints to several extensions of the \LCDM{} model are affected by shifts in the TE PE parameters (see e.g.~Fig.~77 of Ref.~\cite{Planck:2019nip}).\footnote{We note that for \textit{Planck} there exist other likelihood codes which may be used. In this paper we used the Plik likelihood, which is the baseline  \textit{Planck} likelihood for the final third data release (PR3) of the \textit{Planck} collaboration. Another \textit{Planck} likelihood based on PR3 is CamSpec \cite{Aghanim:2018eyx, camspec}, which gives 0.5$\sigma$ shifts relative to Plik in some extensions of $\Lambda$CDM for the TTTEEE data combination. These shifts are due to differences in the treatment of polarization data (Plik and CamSpec provide the same results in TT), which are mostly driven by different choices of polarization efficiencies (see Section 2.2.5 of \cite{Aghanim:2018eyx}). Thus, applying different efficiencies to the Plik likelihood (as done in this paper) provides an accurate proxy of the uncertainty introduced by the difference between the two likelihoods. Moreover, outside of the \textit{Planck} collaboration, new likelihoods (Camspec \cite{2022arXiv220510869R},  Hillipop -- \url{https://github.com/planck-npipe/hillipop}) have recently been proposed based on a new release of \textit{Planck} maps, NPIPE~\cite{NPIPE}. However, while the results are consistent with the ones from PR3, a detailed understanding of differences between data releases and likelihoods is outside of the scope of this paper.}

Another potential systematic effect in the \textit{Planck} data that has to be considered in beyond-\LCDM{} models whose parameters are strongly correlated with the scalar spectral index, $n_s$, involves the choice made for the galactic dust contamination amplitudes \cite{Planck:2019nip}. For the latter, the standard analysis fixes the EE polarization dust amplitudes to values determined by analyzing the 353 GHz map, while the TE dust amplitudes are subject to Gaussian priors (see Fig.~40 of the reference). Lifting such choices does not have significant effects on the parameter estimation (see again Fig. 77 of Ref. \cite{Planck:2019nip}), however, since $f_{\rm EDE}(z_c)$ is strongly correlated with $n_s$ (as shown in  Fig.~\ref{fig: MCMC_full}), we test whether relaxing the dust priors may have a significant impact on our constraints to EDE.
\begin{table*}
\def\arraystretch{1.2}
 \scalebox{1.0}{
 \begin{tabular}{|l|c|c|c|} 
    \hline Parameter & \textit{Planck} TT650TEEE & \textit{Planck} TT650TEEE & \textit{Planck} TTTEEE\\
    & +\ACT+SPT-3G& +\ACT+SPT-3G & +\ACT+SPT-3G\\
    & +BAO+Pantheon&+$\phi\phi$+BAO/$f\sigma_8$+Pantheon&+BAO+Pantheon \\
    \hline \hline
    $f_{\rm EDE}(z_c)$  & $0.148(0.163)_{-0.035}^{+0.039}$ & $0.106(0.143)_{-0.044}^{+0.063}$ & $<0.128 (0.100)$\\
    $\log_{10}(z_c)$& $3.524(3.529)_{-0.026}^{+0.028}$  &  $3.494(3.515)_{-0.032}^{+0.083}$ & $3.511(-3.534)_{-0.11}^{+0.12}$\\
    $\theta_i$ & $2.75(2.757)_{-0.065}^{+0.071}$&$2.512(2.743)_{-0.066}^{+0.41}$  & $2.42(2.77)^{+0.62}_{+0.098}$ \\
    \hline
    $H_0$ [km/s/Mpc] & $73.03(73.51)_{-1.5}^{+1.4}$  & $71.45(72.53)_{-1.7}^{+2.1}$  &$69.72( 70.78)^{+1.1}_{-1.8}$\\
    $100~\omega_b$& $2.273(2.272)_{-0.018}^{+0.016}$ & $2.268(2.261)_{-0.02}^{+0.017}$ & $2.254(2.254)\pm 0.016$\\
    $\omega_{\rm cdm}$& $0.1349(0.1368)\pm0.005$ &$0.1303(0.1345)_{-0.0058}^{+0.0068}$ & $0.1256(0.1299)_{-0.0056}^{+0.0038}$\\
    $10^9A_s$ &  $2.136(2.138)_{-0.038}^{+0.034}$& $2.129(2.155)_{-0.034}^{+0.033}$  & $2.130(2.135)\pm0.038$\\
    $n_s$& $0.9965(0.9977)_{-0.0077}^{+0.0075}$ & $0.9899(0.9931)_{-0.0076}^{+0.0092}$  & $0.9804(0.9846)\pm 0.0075$ \\
    $\tau_{\rm reio}$& $0.0505(0.0498)_{-0.0075}^{+0.0078}$ & $0.0516(0.0549)_{-0.0073}^{+0.0071}$   & $0.0546(0.0521)\pm0.0073$\\
    \hline
    $S_8$ & $0.838(0.841)\pm0.015$& $0.836(0.845)\pm0.014$  &$0.835(0.842)\pm0.014$\\
    $\Omega_m$ &$0.297(0.297)_{-0.006}^{+0.007}$ & $0.301(0.299)_{-0.007}^{+0.006}$    &$0.306(0.306)\pm0.006$ \\
    Age [Gyrs] &$12.95(12.86)_{-0.23}^{+0.22}$ & $13.18(12.99)_{-0.33}^{+0.26}$  &$13.45(13.24)_{-0.16}^{+0.31}$  \\
    \hline
    $\Delta \chi^2_{\rm min}({\rm EDE}-\Lambda{\rm CDM})$ & -14.4& -11.4 & -9.4\\
     Preference over $\Lambda$CDM & 99.8\%  ($3.0\sigma$) &  99.0\% ($2.6\sigma$)&97.6\% ($2.3\sigma$) \\
    \hline
\end{tabular} }
\caption{The mean (best-fit) $\pm 1\sigma$ errors of the cosmological parameters reconstructed from analyses of various data sets (see column title) in the EDE model when including data beyond \textit{Planck} TT650TEEE+\ACT{}+SPT-3G. For each data set, we also report the best-fit $\chi^2$ and improvement $\Delta\chi^2\equiv\chi^2({\rm EDE})-\chi^2(\Lambda{\rm CDM})$.}
\label{tab:non-CMB}
\end{table*}

\begin{figure*}
    \centering
    \includegraphics[width=1.35\columnwidth]{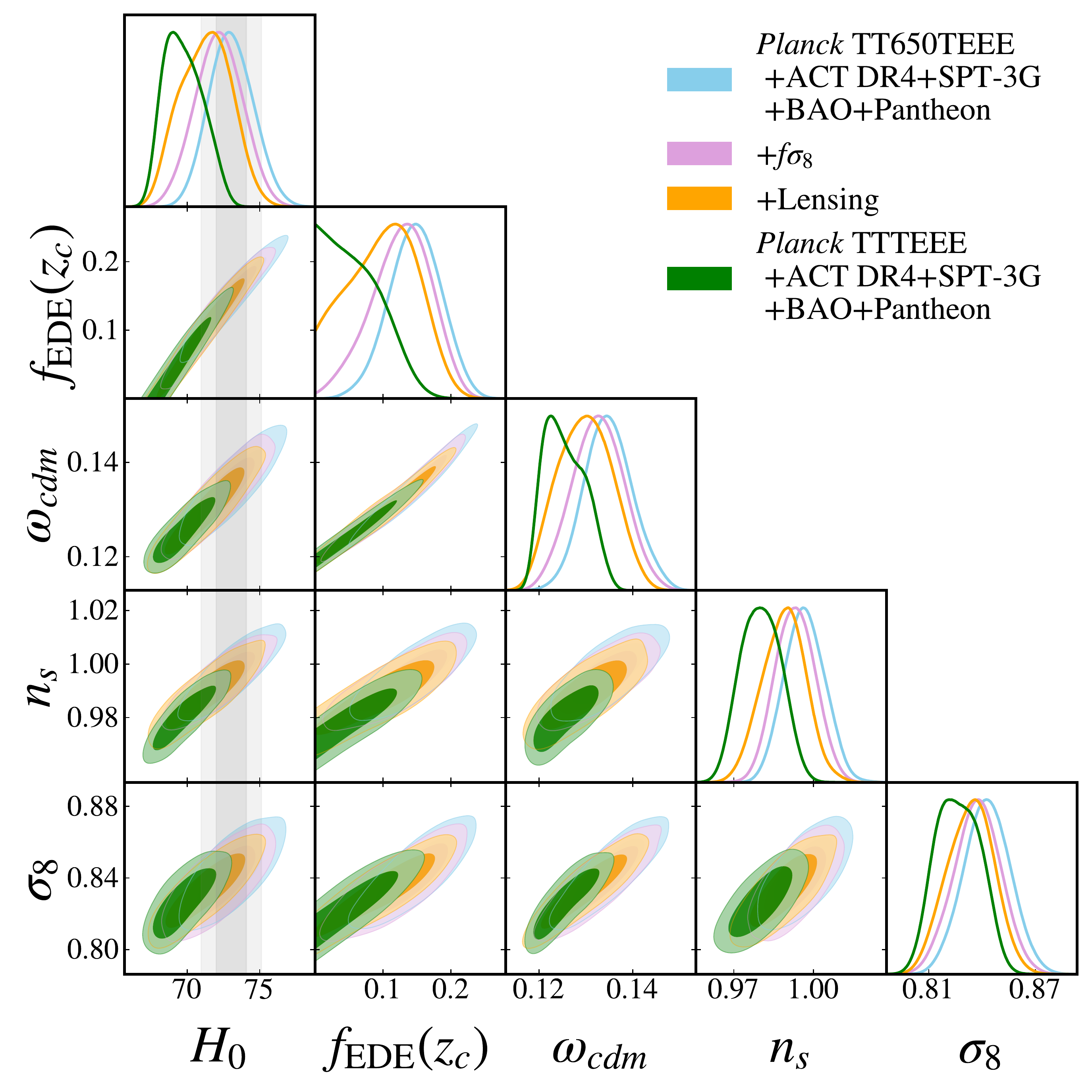}
    \caption{1D and 2D posterior distributions (68\% and 95\% CL) for a subset of the cosmological parameters for different data set combinations fit to EDE. The vertical gray band represents the $H_0$ value reported by the \SHOES{} collaboration \cite{Riess:2021jrx}, $H_0=73.04 \pm 1.04$ km/s/Mpc. The non-CMB data tend to prefer lower values of $n_s$ and $\omega_{\rm cdm}$ leading to lower values of $f_{\rm EDE}(z_c)$. The overall preference for EDE is relatively unchanged when including the BAO and SNe data. Including the full \textit{Planck} data leads to a value of $f_{\rm EDE}(z_c)$ consistent with zero at $\sim 1 \sigma$.}
    \label{fig:external}
\end{figure*}

In order to test the robustness of our results against these possible known sources of systematics, we perform two additional fits of EDE to {\it Planck} TT650TEEE data: one in which we fix the PE calibration factors to the values reported in Eq.~(47) of Ref. \cite{Planck:2019nip}, and another where we place uniform priors on six additional nuisance parameters describing the dust contamination amplitudes in the EE power spectrum. The results of this analysis are shown in Fig.~\ref{fig:syst}, presented in Appendix \ref{app:syst}. We find that the \textit{Planck} preference for EDE vanishes when the TE PE parameters are fixed to the non-standard values ($\Delta\chi^2=-5.1$). Interestingly, the ACT collaboration also found that a potential systematic error in their TE spectra can reduce the preference for EDE within \ACT{} data~\cite{Hill:2021yec}, although not quite as drastically as we find here for {\it Planck}. On the other hand, allowing the dust contamination amplitudes in EE to vary freely has only a marginal effect on the preference for EDE ($\Delta\chi^2=-10.2$). 

\subsection{Impact of non-CMB data}
In Fig.~\ref{fig:external} we show the 1D and 2D posteriors for a subset of the cosmological parameters when including: (i) probes of the late-time expansion history, namely BAO and the (uncalibrated) Pantheon SNe, and (ii) probes of the clustering of matter at late times, namely  $f\sigma_8$ and  \textit{Planck} lensing. A complete list of constraints is given in Table~\ref{tab:non-CMB}.

The inclusion of BAO and Pantheon SNe has a relatively small effect on the preference for EDE over $\Lambda$CDM, slightly reducing it to $3.0\sigma$ ($\Delta\chi^2 = -14.4$). On the other hand, when both $f\sigma_8$ and the \textit{Planck} lensing power spectrum are included, the preference for EDE over \LCDM{} is reduced to $2.6 \sigma$ ($\Delta\chi^2 = -11.4$). It is well known that EDE cosmologies can be tested using measurements of the clustering of matter, since their preferred values of $\omega_{\rm cdm}$ and $n_s$ predict larger clustering at small scales than \LCDM{} \cite{Hill:2020osr,DAmico:2020ods,Ivanov:2020ril,Murgia:2020ryi,Klypin:2020tud}. In fact, the value of $S_8 = 0.829_{-0.019}^{+0.017}$ reconstructed in the EDE cosmology from {\em Planck} TT650TEEE+SPT-3G+\ACT{} is in slight tension\footnote{The level of tension is in fact smaller than in the fiducial {\em Planck} $\Lambda$CDM cosmology \cite{Aghanim:2018eyx,Heymans:2020gsg,DES:2021wwk}, but it is slightly larger than in the $\Lambda$CDM cosmology extracted from {\em Planck} TT650TEEE+SPT-3G+\ACT, see Tab.~\ref{tab:full}.} with the $S_8$ measurements from KiDS-1000+BOSS+2dfLenS \cite{Heymans:2020gsg} ($2.3\sigma$), and DES-Y3 \cite{DES:2021wwk} ($2.1\sigma$). Therefore, it is not surprising that probes of the clustering of matter at late times have a more significant impact on the EDE fit. This is evident in Fig.~\ref{fig:external}: both $f\sigma_8$ and estimates of the lensing potential power spectrum prefer lower values of $n_s$ and $\omega_{\rm cdm}$, leading to a decrease in the marginalized values of $f_{\rm EDE}(z_c)$. However, it is interesting to note that the resulting posterior distribution for the Hubble constant shifts to $H_0 = 71.45 ^{+2.1}_{-1.7}$ km/s/Mpc, i.e. with a central value still significantly higher than in \LCDM{}. Stronger constraints on EDE may be obtained from analyses making use of the full shape of BOSS DR12 data\footnote{ Although these constraints are debated \cite{Murgia:2020ryi,Klypin:2020tud,Niedermann:2020qbw,Smith:2020rxx} and a recently raised potential issue with the calibration of the window function may affect such constraints \cite{Beutler:2021eqq}.}  \cite{DAmico:2020ods,Ivanov:2020ril} or from including additional surveys such as KiDS-1000~\cite{KiDS:2020suj}, DES-Y3 \cite{DES:2021wwk} and HSC \cite{HSC:2018mrq}. A fully satisfactory resolution of the `$S_8$ tension', if not due to systematic errors, e.g. from galaxy assembly bias and baryonic effects \cite{Amon:2022ycy}, may require a more complicated EDE dynamics \cite{Karwal:2021vpk,McDonough:2021pdg,Sabla:2022xzj} or an independent mechanism \cite{Jedamzik:2020zmd,Allali:2021azp,Clark:2021hlo}. 

Finally, in Appendix~\ref{app:ext_full} we present results of combined analyses with a prior on $H_0$ as reported by \SHOES{} \cite{Riess:2021jrx}. We find that when considering the combination of {\it Planck} TT650TEEE+ACT DR4+SPT-3G+ BAO+Pantheon+\SHOES{} the EDE model is favored at 5.3$\sigma$ over $\Lambda$CDM, with $f_{\rm EDE}(z_c)=0.143_{-0.026}^{+0.023}$ and $H_0=72.81_{-0.98}^{+0.82}$ km/s/Mpc. The inclusion of the full \textit{Planck} TT power spectrum, lensing power spectrum and $f\sigma_8$ measurement reduces the preference to $4.3\sigma$, but the EDE model still provides an excellent fit to all data sets, and a potential resolution to the `Hubble tension'.

\subsection{Impact of \textit{Planck} high-$\ell$ TT data} 
In Fig.~\ref{fig:external} we show the parameter reconstructed posteriors when including the full range of the \textit{Planck} TT power spectrum. In that case, we find that the EDE contribution is constrained to be at most $f_{\rm EDE}(z_c)<0.128$ (95\% CL) with a corresponding $H_0=69.7_{-1.8}^{+1.1}$ km/s/Mpc (see Table \ref{tab:non-CMB}), while the preference for EDE drops to the $2.3\sigma$ level (with a best fit value $f_{\rm EDE}(z_c)= 0.1$). We note that, although the  posterior distribution of $f_{\rm EDE}(z_c)$ is compatible with zero at $1\sigma$, it is interesting that the preference, computed using the $\Delta \chi^2$ statistics with three degrees of freedom\footnote{This likely indicates that the true significance of the preference over $\Lambda$CDM is lower than the one reported here, similarly to the way with which {\em local} and {\em global} significance can differ.}, stays above the $2\sigma$ level. This is reminiscent of the difference between the results reported using an EDE model with only one parameter \cite{Murgia:2020ryi,Smith:2020rxx,Niedermann:2020dwg}, or using a frequentist approach through a profile likelihood analysis \cite{Herold:2021ksg}, which led to a $2.2\sigma$ preference for EDE from full {\em Planck} data, as opposed to MCMC analyses that only find upper limits on $f_{\rm EDE}(z_c)$ \cite{Hill:2020osr,Murgia:2020ryi}. In addition to this, the marginalized constraints on $f_{\rm EDE}(z_c)$ using {\em Planck} TTTEEEE with \ACT{} and SPT-3G are roughly $50\%$ {\em weaker} than constraints from {\em Planck} only. 

\begin{figure*}
    \centering
    \includegraphics[width=0.8\linewidth]{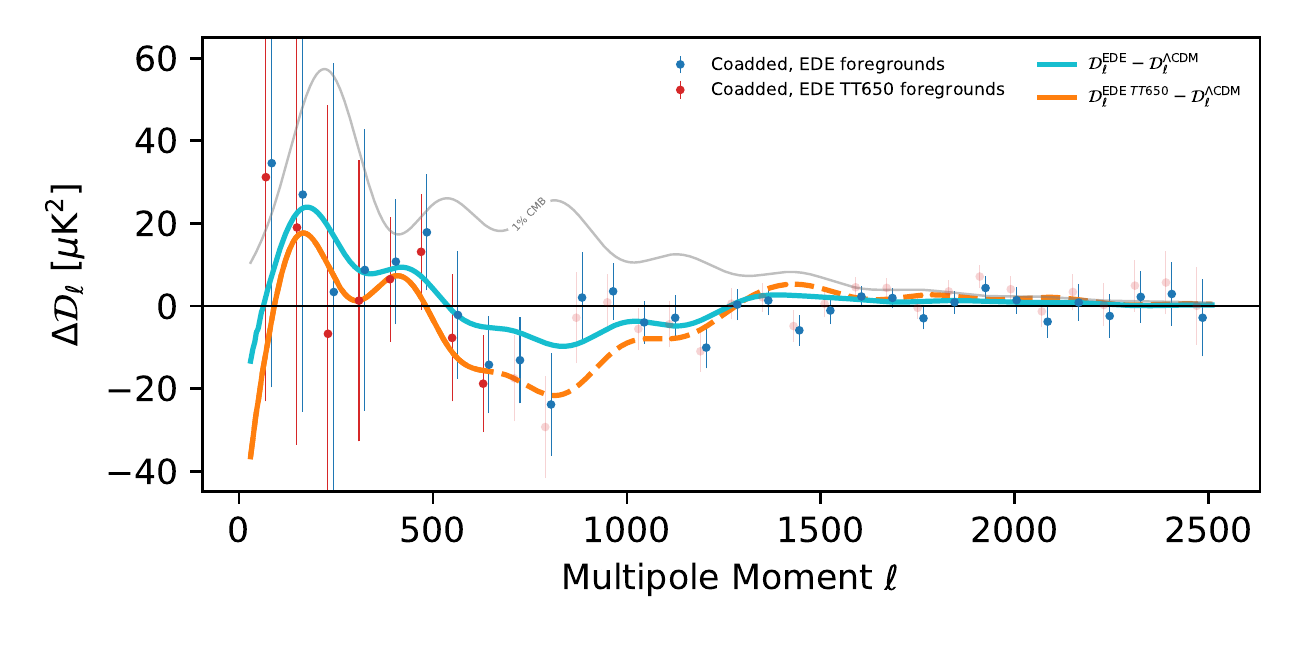}
    \caption{Residual plot of the \textit{Planck} TT data with respect to the reference \LCDM{} best-fit model for the \textit{Planck} TTTEEE+\ACT{}+SPT-3G data set combination. The orange line corresponds to the difference between the EDE best-fit model to the data combination \textit{Planck} TT650TEEE+\ACT{}+SPT-3G (`EDE TT650' in the legend) and the reference \LCDM{} model. The blue line is the same for full \textit{Planck} TTTEEE+\ACT{}+SPT-3G (`EDE' in the legend). Coadded data residuals are computed with respect to the reference \LCDM{} cosmological model but using the best-fit nuisance parameters for each of the two EDE cases (TT650 for the red points and full TT for the blue ones). Since in the TT650 case the high-$\ell$ data, shown in red transparent data points, do not enter the parameter determination, high-$\ell$ foreground parameters are not determined. Therefore, in this case they have been obtained by minimizing the \textit{Planck} TT likelihood when fixing the $C_\ell$ and low-$\ell$ nuisances to the \textit{Planck} TT650TEEE+\ACT{}+SPT-3G best-fit model. At $\ell>900$, the red transparent residual data points are very close to the blue ones, which indicates that the difference in nuisance parameters between the two cases is small. The high-$\ell$ orange best-fit line predicted by the EDE TT650 case is far from the residual data points, regardless of the nuisance model chosen. It is therefore the high-$\ell$ TT data which drives the best-fit closer to \LCDM{}, from the orange line toward the blue one.}
    \label{fig: EDE_plot_final}
\end{figure*}

Given that the posterior distribution of $f_{\rm EDE}(z_c)$ is compatible with zero at $1\sigma$, we conservatively interpret these results as an indication that the full \textit{Planck} TT power spectrum slightly disfavors the EDE cosmology preferred by the other data sets. We leave a more robust determination of this (in)consistency to future work.

\begin{figure}
    \centering
    \includegraphics[width=0.95\columnwidth]{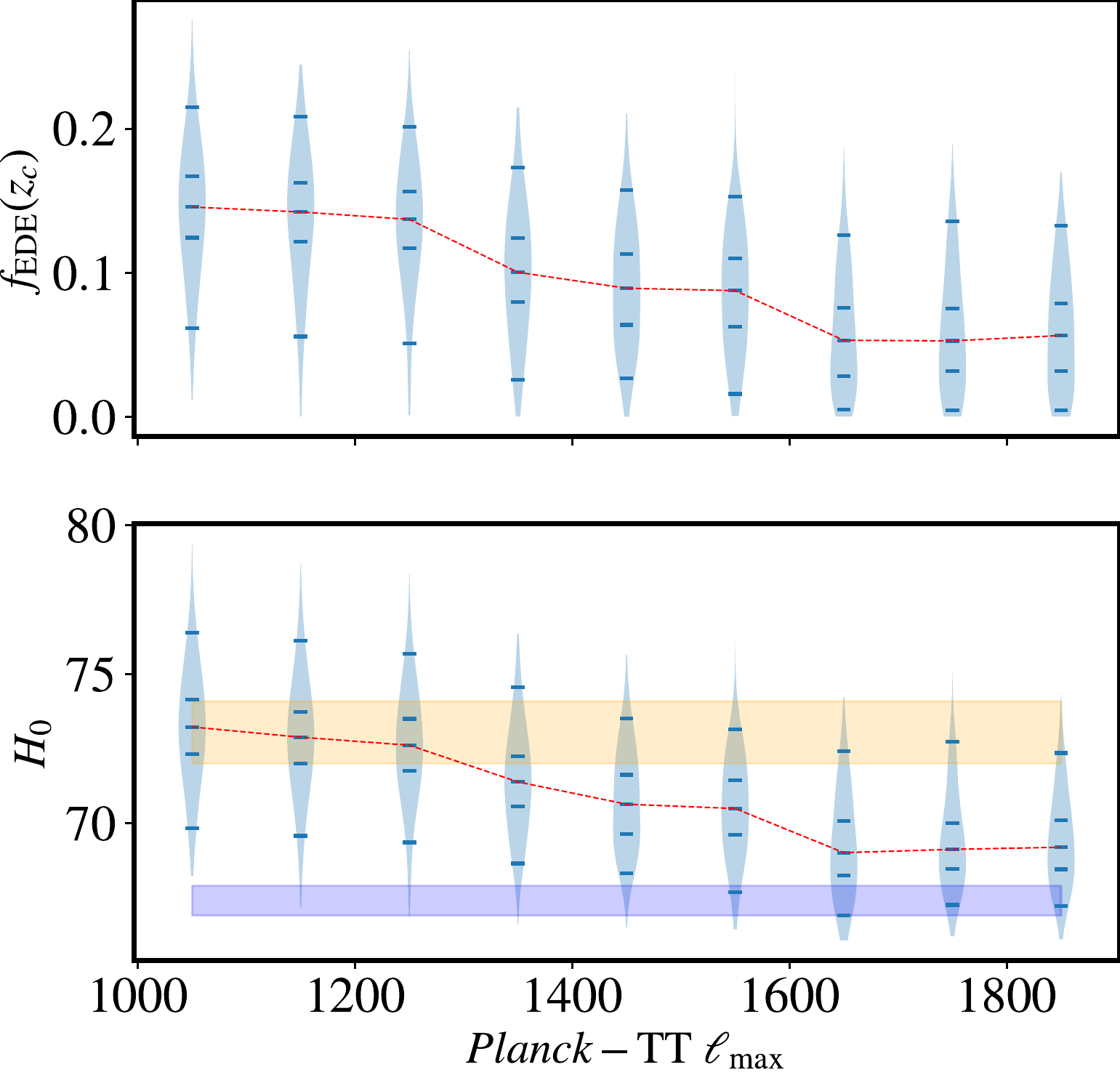}
    \caption{The posterior distribution for $f_{\rm EDE}(z_c)$ and $H_0$ as a function of the maximum TT multipole for \textit{Planck} TT($\ell_{\rm max}$)TEEE+\ACT. The yellow and purple bands in the bottom panel give the \SHOES\ and the full \textit{Planck} values for $H_0$, respectively. Note that, following Ref.~\cite{Aiola:2020azj}, in the chains used to make this figure we restricted the \ACT{} temperature bins so as to remove any overlap with \textit{Planck} up until $\ell_{\rm max} = 1800$. As the \textit{Planck} TT $\ell_{\rm max}$ is increased the preference for a non-zero contribution of EDE is decreased, leading to a smaller inferred value of $H_0$.}
    \label{fig:TT_ellmax}
\end{figure}

We show in Fig.~\ref{fig: EDE_plot_final} the difference between the temperature power spectra obtained in the EDE best-fit to {\it Planck} TT650TEEE+\ACT{}+SPT-3G or full {\it Planck} TTTEEE+\ACT{}+SPT-3G, and the $\Lambda$CDM fit to full {\it Planck} TTTEEE+\ACT{}+SPT-3G. We also show {\it Planck} TT data residuals with respect to the $\Lambda$CDM model. To gauge the role of foregrounds in affecting the preference for EDE, we compare the data residuals for the foreground models obtained from the restricted fit to those obtained in the fit to the full range of data. One can see that data residuals are fairly similar, indicating that high-$\ell$ foregrounds are not strongly correlated with EDE, and cannot be the reason for which \textit{Planck} high-$\ell$ TT data seems to disfavor EDE. Additionally, one can see that data points up to $\ell \sim 850$ are in good agreement with the EDE best-fit model, but start diverging around $\ell \sim 900$.

To better understand the impact of the \textit{Planck} TT power spectrum, in Fig.~\ref{fig:TT_ellmax} we show how the preference for EDE evolves as we increase the considered range of the \textit{Planck} TT power spectrum in steps of $\Delta \ell=100$.\footnote{Here we do not include the SPT-3G data for sake of computational speed, but we have explicitly checked with a few dedicated runs that its addition does not impact our conclusions.} The evidence for EDE over \LCDM{} (and the corresponding increased value of $H_0$) starts to drop off once the TT multipoles $\ell \gtrsim 1300$ are included. This is consistent with the fact that \textit{Planck} gains most of its statistical power between $\ell \sim 1300$ and $\ell \sim 2000$, and drives the model to be extremely close to $\Lambda$CDM. Given that high-$\ell$ \ACT{} temperature power spectrum is partly driving the preference for EDE, as mentioned previously, this may hint to a small inconsistency between {\it Planck} and \ACT{} temperature data (see also Ref.~\cite{Handley:2019wlz}), although at the current level of significance a statistical fluctuation cannot be ruled out.

\section{Summary and conclusions}\label{sec:concl}
We have found that when analyzing EDE using \ACT, SPT-3G, and \textit{Planck} measurements of the CMB a consistent story emerges if we exclude the \textit{Planck} temperature power spectrum at high-$\ell$: an EDE component consisting of $\sim 10-15\%$ of the total energy density at a redshift $\log_{10}(z_c) \simeq 3.5$ with an initial field displacement of $\theta_i \simeq 2.7$ and a corresponding increase in the inferred value of the Hubble constant with $H_0 \simeq 73-74$ km/s/Mpc, in contrast to \LCDM{} which gives $H_0 \simeq 68$ km/s/Mpc (see Table \ref{tab:CMB}). 

Such hints for an EDE cosmology are present when combining \textit{Planck} polarization power spectra with \textit{Planck} TT excised at $\ell>650$ (2.2$\sigma$), and when adding ACT DR4 (3.3$\sigma$) or SPT-3G (2.4$\sigma$). Combining all three CMB data sets  yields a 3.3$\sigma$ preference for EDE over \LCDM{}. The inclusion of the \textit{Planck} polarization data effectively removes the differences between the best-fits of the measurements of the lowest polarization multipoles by \ACT{} and SPT-3G, and emphasizes the new information that these observations provide. Indeed, together with \textit{Planck} polarization data the EDE best-fits for both \ACT{} and SPT-3G visually come into closer agreement (although a more careful analysis of their consistency is left for future work). This preference remains at the 3$\sigma$ level when adding the Pantheon SNe and the BAO standard ruler, increases above 5$\sigma$ when including an $H_0$ prior from \SHOES{}, and is mildly reduced when considering CMB lensing potential data or estimates of~$f\sigma_8$. 

We find that these results remain unchanged when increasing the maximum \textit{Planck} TT multipole until ${\ell=1300}$. On the contrary, the inclusion of small angular scale data from the \textit{Planck} temperature power spectrum above that multipole decreases this preference to $2.3 \sigma$ (in the absence of a $H_0$ prior). This is consistent with the fact that \textit{Planck} high-$\ell$ TT data have most of their constraining power at those scales, and drive parameters very close to their $\Lambda$CDM values, limiting the ability to exploit degeneracies between $\Lambda$CDM and EDE parameters. There have been several previous studies looking into the consistency between the `low' ($\ell \lesssim 1000$) and `high' TT multipoles  (see e.g. Refs.~\cite{Addison:2015wyg,Planck:2018vyg,Planck:2016tof}). The high-$\ell$ TT power spectrum has a slight ($\sim 2 \sigma$) preference for higher $\omega_{\rm cdm}$, higher amplitude ($A_s e^{-2\tau_{\rm reio}}$), and lower $H_0$. However, an exhaustive exploration of these shifts indicates that they are all consistent with expected statistical fluctuations~\cite{Planck:2016tof}. Although there may be localized features in the high-$\ell$ TT power spectrum which are due to improperly modeled foregrounds (see Sec.~6.1 in Ref.~\cite{Planck:2018vyg}), under the assumption of \LCDM{} there is no evidence that these data are broadly biased. However, it is interesting to note that the \ACT{} TT data at these multipoles are consistent with the preference for EDE.

Moreover, it is well known that \textit{Planck} polarization data may suffer from some systematic uncertainties which may, in turn, impact our conclusions. The most significant potential source of systematics would imply a change in the TE polarization efficiencies. We explore this by re-analyzing the \textit{Planck} constraints on EDE using different TE polarization efficiencies and find that the \textit{Planck} TT650TEEE preference for EDE largely reduces. A similar analysis conducted in Ref.~\citep{Hill:2021yec} accounted for a possible unknown source of systematics in \ACT{} TE data and showed that it also reduces the \ACT's preference for EDE. When allowing the EE dust amplitudes to vary we found almost no change to the constraints on EDE.

It is thus clear that future, high-precision, CMB temperature and polarization data will be necessary to disentangle whether the reported preference for EDE over \LCDM{} is driven by systematics or a hint of new physics (or, possibly, a statistical fluctuation). In particular, the precision expected from upcoming data releases from SPT and ACT (as mentioned in the conclusions of Refs.~\cite{SPT-3G:2021eoc, SPT:2021slg, ACT:2020frw, ACT:2020gnv}) with combined temperature, polarization, and lensing likelihoods will be capable of constraining the parameter space of the EDE model even more tightly\footnote{In the future, CMB spectral distortions will also be able to determine the value of $n_s$ with a high significance, thereby testing the EDE ability to address the $H_0$ tension independently of CMB anisotropy data~\cite{Lucca:2020fgp}.} as well as of clarifying how the small-scale CMB TT measurements impact the EDE constraints.

This will not only provide a valuable cross-check on the \textit{Planck} measurements, but also an opportunity to obtain tight and robust constraints through joint analyses, which can be of primary importance to test physics scenarios beyond $\Lambda$CDM with CMB data (as in the case of e.g. primordial magnetic fields \cite{Jedamzik:2020zmd,Galli:2021mxk}, sterile neutrino self-interactions \cite{Corona:2021qxl} and New EDE \cite{Niedermann:2019olb, Niedermann:2020dwg}) as our work demonstrates. In fact, based on the analyses previously conducted in \citep{Poulin:2021bjr, Schoneberg:2021qvd}, we also carried out preliminary tests to check whether the same data set combinations that lead to a preference for EDE would also display a similar behavior in other beyond-$\Lambda$CDM models (such as New EDE and varying electron mass), finding that this is not the case. The same conclusion was also recently reached in \cite{Schoneberg:2022grr} in the context of the Wess Zumino Dark Radiation model introduced in \cite{Aloni:2021eaq}. Although a more in-depth analysis is left for future work, this is already indicative of the very important role that future data might play in testing and distinguishing non-standard cosmological models.

\acknowledgments
The authors thank J. Colin Hill, Thibaut Louis and Adam G. Riess for useful comments and suggestions. This work used the Strelka Computing Cluster, which is run by Swarthmore College. TLS is supported by NSF Grant No.~2009377, NASA Grant No.~80NSSC18K0728, and the Research Corporation. ML is supported by an F.R.S.-FNRS fellowship, by the `Probing dark matter with neutrinos' ULB-ARC convention and by the IISN convention 4.4503.15. LB acknowledges support from the University of Melbourne and the Australian Research Council (DP210102386). SG is supported by the European Research Council (ERC) under the European Union’s Horizon 2020 research and innovation program (grant agreement No 101001897). This work has been partly supported by the CNRS-IN2P3 grant Dark21. The authors acknowledge the use of computational resources from the Excellence Initiative of Aix-Marseille University (A*MIDEX) of the “Investissements d’Avenir” programme. This project has received support from the European Union’s Horizon 2020 research and innovation program under the Marie Skodowska-Curie grant agreement No 860881-HIDDeN.

\newpage

\bibliography{bibliography}{}
\newpage
\newpage

\onecolumngrid
\appendix

\section{Supplementary material on the CMB constraints}\label{app: tables}
\begin{figure*}[h!]
     \centering
     \includegraphics[width=0.95\linewidth]{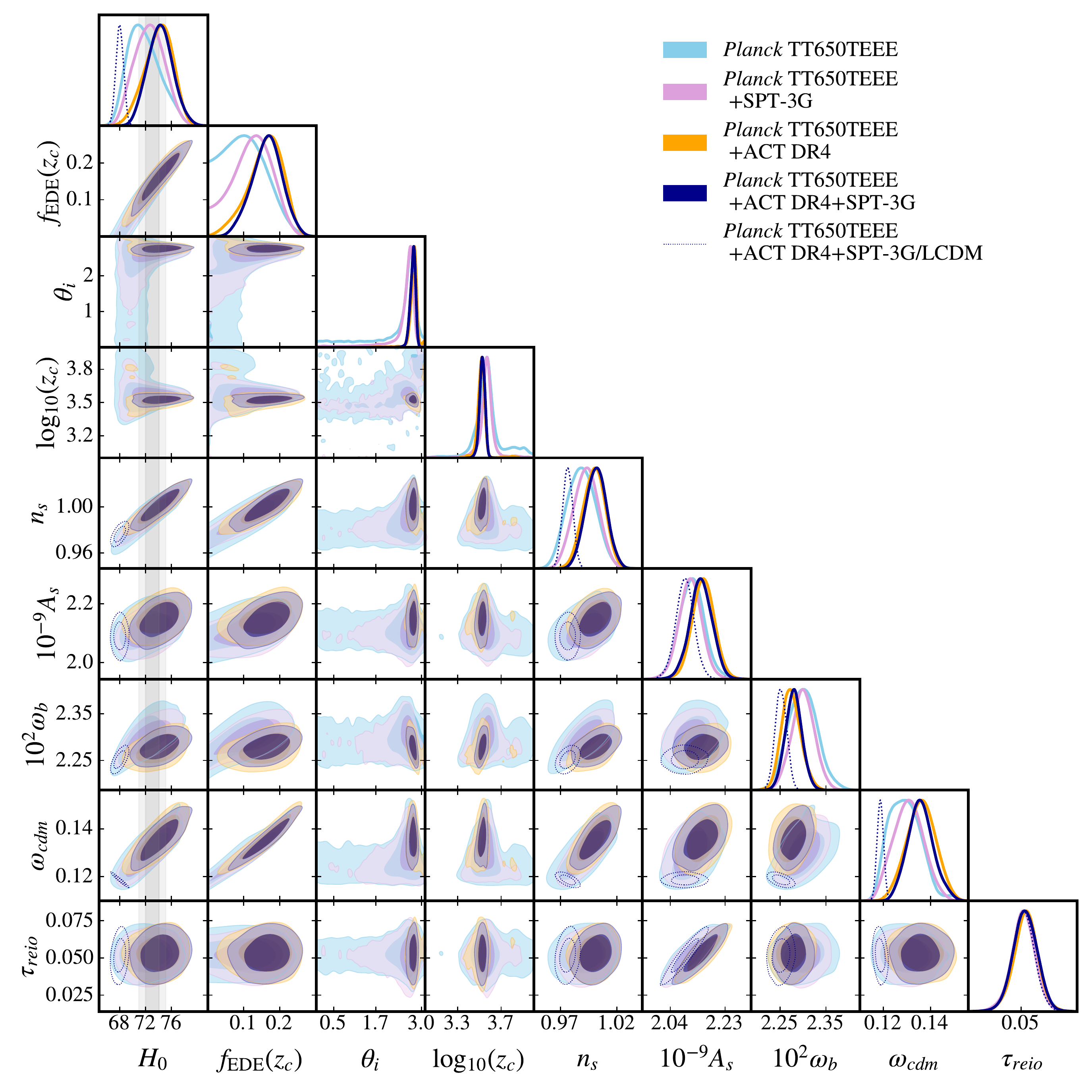}
     \caption{A triangle plot displaying the posterior distributions of the full set of cosmological parameters for the same data set combinations shown in Fig.~\ref{fig: MCMC_res}.}
     \label{fig: MCMC_full}
\end{figure*}

In this appendix we provide constraints on the EDE model for different combinations of CMB probes (Fig.~\ref{fig: MCMC_full} and Table~\ref{tab:CMB}) to be compared with those already presented in Table \ref{tab:full} for the full combination \ACT{}+SPT-3G+\textit{Planck} TT650TEEE (which we repeat in the right column of Table \ref{tab:CMB} for convenience). Additional discussion about the behaviours of the single parameters can be found in e.g. \citep{Poulin:2018dzj, Poulin:2018cxd,  Smith:2019ihp, Hill:2020osr}.

\begin{table*}[h!]
 \scalebox{1}{
 \begin{tabular}{|l|c|c|c|c|} 
    \hline  Parameter &\textit{Planck} TT650TEEE & \textit{Planck} TT650TEEE & \textit{Planck} TT650TEEE & \textit{Planck} TT650TEEE\\
    & & + ACT DR4 & + SPT-3G & + ACT DR4 + SPT-3G\\
    \hline \hline
    $f_{\rm EDE}(z_c)$  & $0.101(0.163)_{-0.073}^{+0.054}$ & $0.162(0.178)_{-0.039}^{+0.06}$&$0.123(0.156)_{-0.049}^{+0.062}$ & $0.163(0.179)_{-0.04}^{+0.047}$   \\
    $\log_{10}(z_c)$& $3.585(3.573)_{-0.13}^{+0.083}$  &  $3.529(3.521)_{-0.049}^{+0.03}$ &$3.566(3.570)_{-0.058}^{+0.062}$ & $3.526(3.528)_{-0.024}^{+0.028}$\\
    $\theta_i$ & $2.262(2.732)_{-0.012}^{+0.73}$  &$2.784(2.806)_{-0.093}^{+0.098}$  & $2.5(2.706)_{-0.048}^{+0.36}$&$2.755(2.777)_{-0.06}^{+0.087}$ \\
    \hline
    $H_0$ [km/s/Mpc]  & $71.74(74.30)_{-3}^{+2}$ &  $74.1(74.8)_{-2.1}^{+2.6}$&$72.58(73.91)_{-2.3}^{+2.3}$ &$74.2(74.83)_{-2.1}^{+1.9}$   \\
    $100~\omega_b$ & $2.299 (2.305) \pm0.033$  & $2.274 (2.271)_{-0.023}^{+0.02}$ & $2.3(2.303)_{-0.025}^{+0.026}$& $2.279(2.278)_{-0.02}^{+0.018}$ \\
    $\omega_{\rm cdm}$& $0.1291(0.1351)_{-0.0081}^{+0.0059}$  & $0.1362(0.1376)_{-0.0065}^{+0.0068}$  & $0.1307(0.1339)_{-0.0068}^{+0.0063}$&$0.1356(0.1372)_{-0.0059}^{+0.0053}$\\
    $10^9A_s$ & $2.116(2.132)_{-0.049}^{+0.043}$ & $2.155(2.159)_{-0.041}^{+0.039}$  &$2.109(2.12)_{-0.041}^{+0.04}$ & $2.145(2.146)_{-0.04}^{+0.041}$\\
    $n_s$ & $0.9886(0.998)\pm0.013$ &  $1(1.0022)\pm0.01$ & $0.9926(0.9978)\pm0.011$ &$1.001(1.003)_{-0.0096}^{+0.0091}$   \\
    $\tau_{\rm reio}$ & $0.0524(0.0534)_{-0.0083}^{+0.0086}$  &$0.0533(0.0538)_{-0.008}^{+0.0081}$ &$0.0513(0.05173)_{-0.008}^{+0.0087}$ &$0.0527(0.052)_{-0.0084}^{+0.0086}$   \\
    \hline
    $S_8$ &  $0.823 (0.822)_{-0.025}^{+0.022}$ &  $0.835 (0.834)\pm0.021$ &$0.818(0.819)\pm0.021$ & $0.829(0.829)_{-0.019}^{+0.017}$\\
    $\Omega_m$ &   $0.297(0.288)\pm0.012$ & $0.291(0.288)_{-0.012}^{+0.011}$  & $0.293(0.288)\pm0.011$&$0.289(0.287)\pm0.009$ \\
    Age [Gyrs] & $13.19 (12.83)_{-0.28}^{+0.45}$ & $12.84 (12.75)_{-0.36}^{+0.26}$ & $13.08(12.89)_{-0.35}^{+0.3}$ &$12.84(12.75)\pm0.27$ \\
    \hline
    $\Delta \chi^2_{\rm min}({\rm EDE}-\Lambda{\rm CDM})$ & -9.4& -16.1 &-10.4& -16.2 \\
    %Preference over $\Lambda$CDM & 97.5\% ($2.2\sigma$) & 99.9\% ($3.3\sigma$)& 98.4\%  (2.4$\sigma$) & 99.9\% ($3.3\sigma$) \\
    Preference over $\Lambda$CDM & $2.2\sigma$ & $3.3\sigma$ & 2.4$\sigma$ & $3.3\sigma$ \\
    \hline
\end{tabular} }
\caption{The mean (best-fit) $\pm 1\sigma$ errors of the cosmological parameters reconstructed from analyses of various data sets (see column title) in the EDE model. For each data set, we also report the best-fit $\chi^2$ and the $\Delta\chi^2\equiv\chi^2({\rm EDE})-\chi^2(\Lambda{\rm CDM})$. }
\label{tab:CMB}
\end{table*}

\section{Supplementary tables of $\chi^2_{\rm min}$ values per experiment}\label{app: chi2}
In this appendix we report a complete breakdown of the best-fit $\chi^2$ per experiment for both the \LCDM{} (Table~\ref{tab:chi2_lcdm}) and EDE (Table~\ref{tab:chi2_ede}) models. In Table~\ref{tab:chi2_nopol} we also focus on the dependence of the best-fit $\chi^2$ values on the exclusion of \textit{Planck} polarization data.

\begin{table*}[h!]
\def\arraystretch{1.2}
\scalebox{1}{
\begin{tabular}{|l|c|c|c|c|c|c|c|}
    \hline
    \multicolumn{8}{|c|}{$\Lambda$CDM} \\
    \hline
    {\emph{Planck}}~high$-\ell$ TT650TEEE &  1839.9&1843.2 & 1841.3& 1842.3 & 1842.8   & 1842.4 & $-$   \\
    {\emph{Planck}}~low$-\ell$ EE & 395.6&395.7 &  395.7& 395.7& 395.7& 396.1 & 395.9\\
    {\emph{Planck}}~low$-\ell$ TT & 22.1& 21.7&  21.8& 21.7 & 21.3& 21.4 & 22.1\\
    ACT DR4\footnote{In the last column, ACT DR4 data are restricted to $\ell >1800$.} & $-$ & 293.8&$-$  & 296.0& 296.4& 296.0 &  242.3\\
    SPT-3G &$-$  & $-$ &517.6 &519.0 &518.4& 523.7& 520.3\\
    Pantheon SN1a &$-$  & $-$ &$-$  & $-$ &1026.8& 1027.0 & 1026.9 \\
    BOSS BAO low$-z$ & $-$ & $-$ &$-$  &$-$  &   1.5 & 1.6 &1.3\\
    BOSS BAO DR12  & $-$ & $-$ & $-$ &$-$  &  3.7& $-$&4.1\\
    BOSS BAO/$f\sigma_8$ DR12  & $-$ & $-$ & $-$ &$-$ &  $-$ & 6.0 &$-$\\
    {\emph{Planck}}~lensing &$-$ &$-$ &$-$ & $-$&$-$ & 9.0 &$-$ \\
    {\emph{Planck}}~high$-\ell$ TTTEEE &  $-$& $-$ &$-$  & $-$ & $-$&  $-$ & 2349.4\\
    \hline
    total $\chi^2_{\rm min}$  & 2257.6& 2554.4 &2776.4 & 3074.7 &  4106.6& 4123.2 &4562.3\\
    \hline
\end{tabular}}
\caption{Best-fit $\chi^2$ per experiment (and total) for $\Lambda$CDM when fit to different data combinations. Each column corresponds to a different data set combination.}
\label{tab:chi2_lcdm}
\end{table*}

\begin{table*}
\def\arraystretch{1.2}
\scalebox{1}{
\begin{tabular}{|l|c|c|c|c|c|c|c|}
    \hline
    \multicolumn{8}{|c|}{EDE} \\
    \hline
    {\emph{Planck}}~high$-\ell$ TT650TEEE & 1831.9& 1837.6&1833.3 &1836.2 &  1836.2 & 1837.8 & $-$ \\
    {\emph{Planck}}~low$-\ell$ EE & 395.8& 395.9&395.8 & 395.8& 395.9& 396.1 & 395.8\\
    {\emph{Planck}}~low$-\ell$ TT & 20.5&20.4 &20.4 & 20.3& 20.6& 20.8 & 21.4\\
    ACT DR4\footnote{In the last column, ACT DR4 data are restricted to $\ell >1800$.} & $-$ & 284.4&$-$  & 288.0& 288.5& 288.1 & 238.6\\
    SPT-3G &$-$  &$-$  & 516.5& 518.2&  518.6& 522.6 & 519.7 \\
    Pantheon SN1a &$-$  &$-$  &$-$  &$-$  &1026.7& 1026.9 &1026.7 \\
    BOSS BAO low$-z$ &$-$  & $-$ &$-$  &$-$  &  2.3& 2.0 &  1.5\\
    BOSS BAO DR12  &  $-$& $-$ &$-$  & $-$ &3.6 & $-$ & 3.7\\
    BOSS BAO/$f\sigma_8$ DR12  & $-$ & $-$ & $-$ &$-$ &  $-$ & 7.1 &$-$\\
    {\emph{Planck}}~lensing &$-$ &$-$ &$-$ & $-$&$-$ & 10.2 &$-$ \\
    {\emph{Planck}}~high$-\ell$ TTTEEE &  $-$& $-$ &$-$  & $-$ & $-$&  $-$ & 2345.5\\
    \hline
    total $\chi^2_{\rm min}$  & 2248.2& 2538.3& 2766.0&3058.5 &  4092.2& 4111.6 & 4552.9\\
    $\Delta \chi^2_{\rm min}({\rm EDE}-\Lambda{\rm CDM})$ & -9.4& -16.1 &-10.4& -16.2& -14.2& -11.6 & -9.4  \\
    \hline
    %Preference over $\Lambda$CDM & \makecell{97.56\% \\ ($2.3\sigma$)} & \makecell{99.89\% \\ ($3.3\sigma$)}& \makecell{ 98.45\% \\ (2.4$\sigma$)} & \makecell{99.90\% \\ ($3.3\sigma$)} & \makecell{99.73\% \\ ($3.0\sigma$)} & \makecell{99.1\%\\($2.6\sigma$)}& \makecell{97.56\% \\ ($2.3\sigma$)} \\
    Preference over $\Lambda$CDM & $2.3\sigma$ & $3.3\sigma$ & 2.4$\sigma$ & $3.3\sigma$ & $3.0\sigma$ & $2.6\sigma$ & $2.3\sigma$ \\
    \hline
\end{tabular}}
\caption{Best-fit $\chi^2$ per experiment (and total) for EDE when fit to different data combinations. We also report the $\Delta \chi^2_{\rm min}\equiv\chi^2_{\rm min}({\rm EDE})-\chi^2_{\rm min}(\Lambda{\rm CDM})$ and the corresponding preference over $\Lambda$CDM, computed assuming the $\Delta\chi^2$ follows a $\chi^2$-distribution with three degrees of freedom.  }
\label{tab:chi2_ede}
\end{table*}

\begin{table*}
\def\arraystretch{1.2}
\scalebox{1}{
\begin{tabular}{|l|c|c|c|c|c|c|}
    \hline
   &  \multicolumn{3}{|c|}{$\Lambda$CDM}& \multicolumn{3}{|c|}{EDE} \\
    \hline
    {\emph{Planck}}~high$-\ell$ TT650 & 250.7&250.7 &251.8 &250.1 &250.2 &249.5\\
   $\tau$& 0.003& 0.1& 0.004&0.003 &0.02 &0.002\\
    ACT DR4 &288.9 & $-$ &290.5 & 272.6& $-$ &281.5\\
    SPT-3G  & $-$ &517.7 & 519.9 &  $-$&  513.0 & 521.3\\
    \hline
    total $\chi^2_{\rm min}$&539.6 & 768.5& 1062.2&  522.7& 763.2&1052.3 \\
    $\Delta \chi^2_{\rm min}({\rm EDE}-\Lambda{\rm CDM})$&  $-$  &  $-$ &  $-$ & -16.9&  -5.3&-9.9 \\
    \hline
    Preference over $\Lambda$CDM & $-$ &$-$& $-$& %\makecell{99.9\%\\(3.4$\sigma$)} &\makecell{84.9\%\\(1.4$\sigma$)} &\makecell{98.1\%\\(2.3$\sigma$)} \\
    3.4$\sigma$ & 1.4$\sigma$ & 2.3$\sigma$ \\
    \hline
\end{tabular}}
\caption{Best-fit $\chi^2$ per experiment (and total) for EDE when fit to different data combinations that do not include {\em Planck} polarization data. We also report the $\Delta \chi^2_{\rm min}\equiv\chi^2_{\rm min}({\rm EDE})-\chi^2_{\rm min}(\Lambda{\rm CDM})$ and the corresponding preference over $\Lambda$CDM, computed assuming the $\Delta\chi^2$ follows a $\chi^2$-distribution with three degrees of freedom.  }
\label{tab:chi2_nopol}
\end{table*}

\FloatBarrier
\newpage

\section{Supplementary material on the role of {\em Planck} polarization data}
\label{app:nopol}
\begin{figure*}
    \centering
    \includegraphics[width=0.7\linewidth]{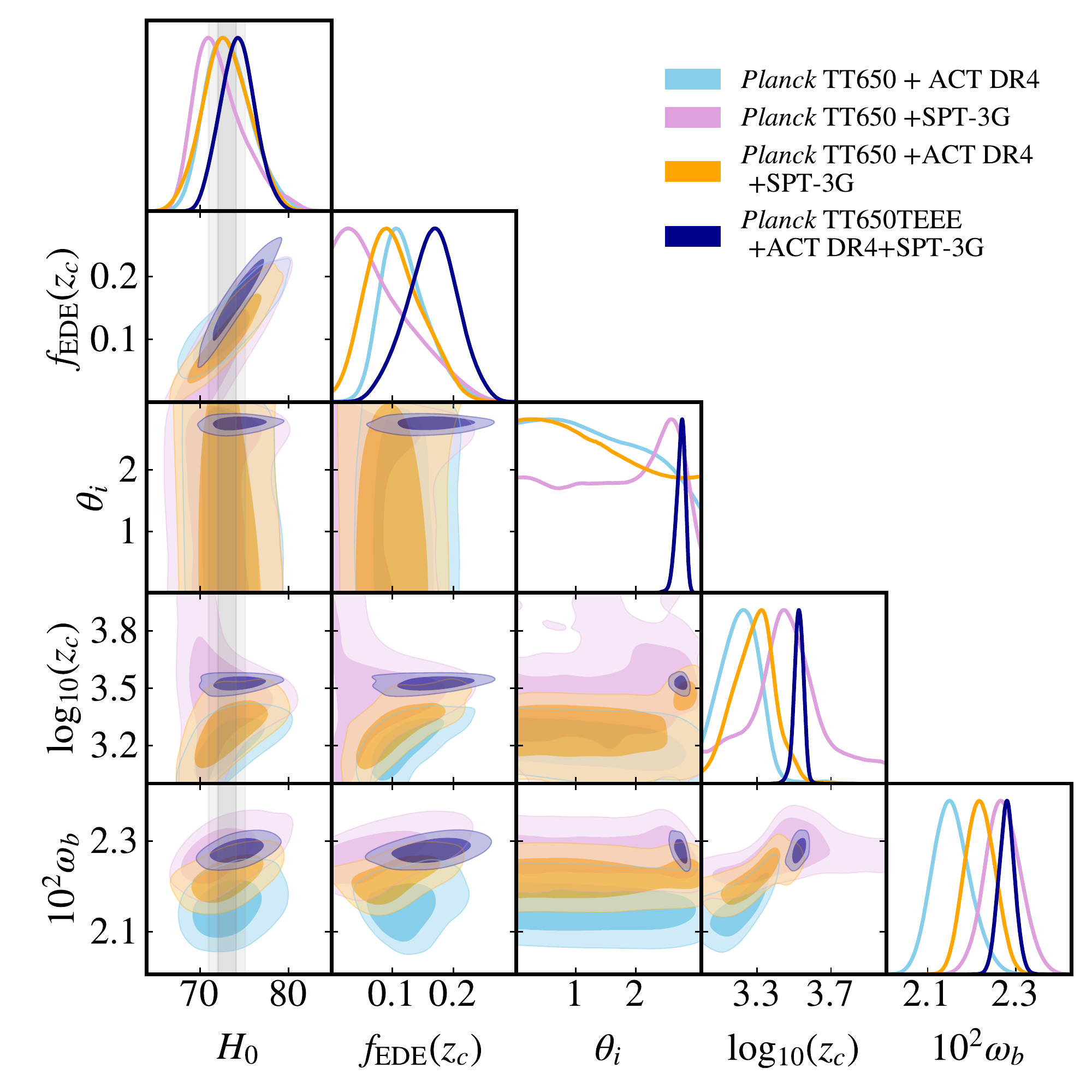}
    \caption{1D and 2D posterior distributions (68\% and 95\% CL) for different data set combinations with and without \textit{Planck} polarization measurements. The vertical gray band represents the $H_0$ value $H_0=73.04 \pm 1.04$ km/s/Mpc as reported by the \SHOES{} collaboration \cite{Riess:2021jrx}. The inclusion of \textit{Planck} polarization significantly narrows the posterior for $\theta_i$ and $\log_{10}(z_c)$.}
    \label{fig:PlanckTT650}
\end{figure*}

\begin{figure*}[h!]
    \centering
     \includegraphics[width=0.7\columnwidth]{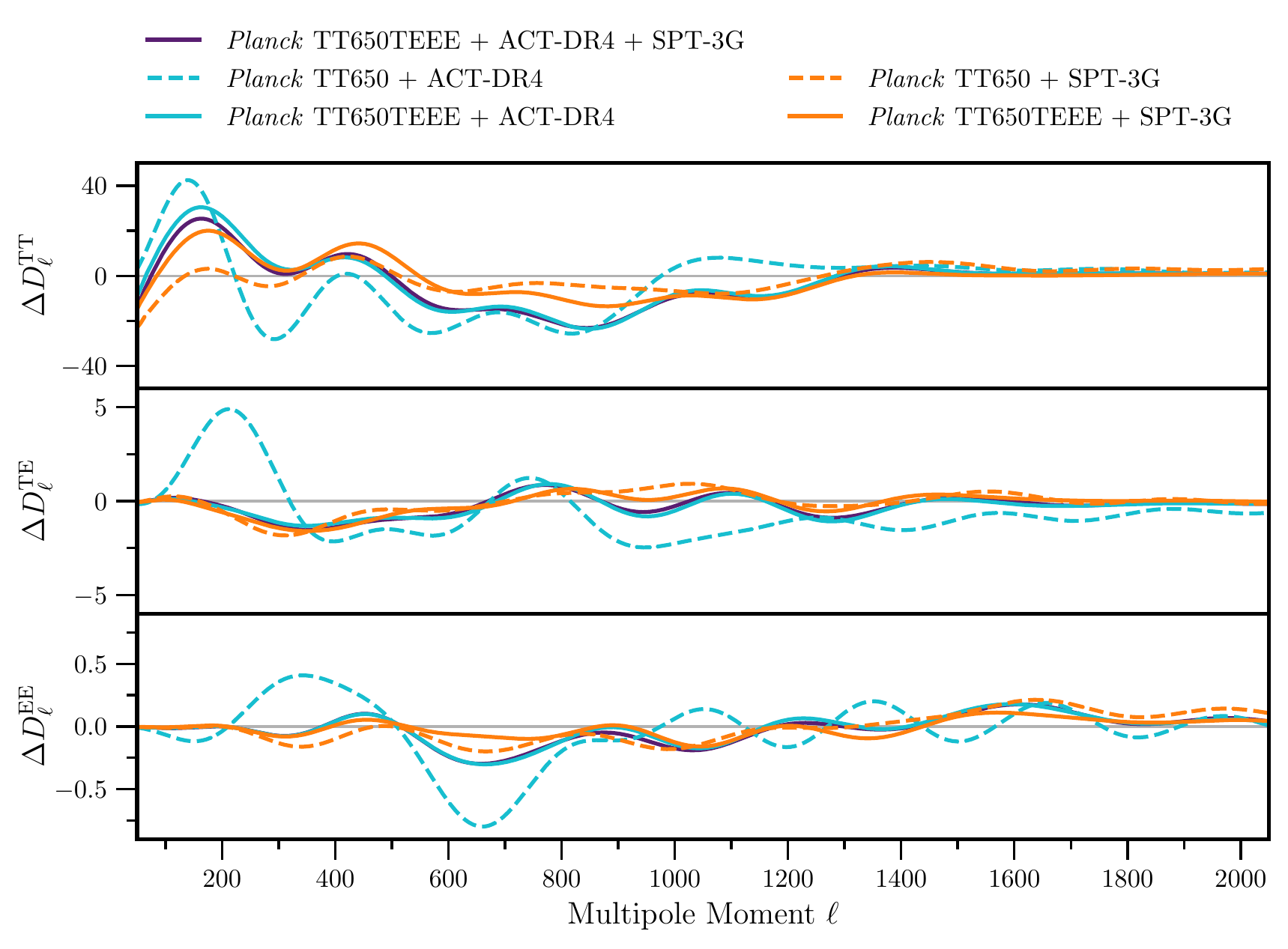}
     \caption{Difference plots (in units of $\mu{\rm K}^2$) of the CMB-only TT, TE and EE power spectra between their respective EDE and \LCDM{} best-fit models for various data set combinations. The addition of \textit{Planck} TEEE data to \textit{Planck} TT650 and either SPT-3G (dashed to solid orange) or ACT DR4 (dashed to solid blue) leads to similar CMB spectra. This indicates that these joint fits favor the same EDE model. The combination of \textit{Planck} TT650TEEE+\ACT{}+SPT-3G is shown in purple.}
     \label{fig: residuals_full}
\end{figure*}

In this appendix we compare results with and without {\em Planck} polarization data. In Table~\ref{tab:TTonly} we present results of analyses performed with \ACT, SPT-3G and \textit{Planck} TT650 data including a prior on the optical depth to reionization as measured by {\em Planck} within $\Lambda$CDM, ${\tau=0.0543\pm0.0073}$ (but no polarization data). A graphical representation of the posterior distributions of the parameters most relevant for our discussion is shown in Fig. \ref{fig:PlanckTT650}.  In that figure, we also include posteriors including \textit{Planck} TEEE data (already presented earlier) to gauge visually the impact of those data on the posteriors. In particular, one can see how the inclusion of the \textit{Planck} polarization data significantly narrows the posterior for $\theta_i$ and $\log_{10}(z_c)$, favoring values of $z_c\sim 10^{3.5}$ and $\theta_i\sim 2.8$. These are slightly larger than the results from analyses combining \ACT{} with {\em Planck} TT650 (although compatible at $\sim2\sigma$), and in good agreement with results from past studies combining {\em Planck} with a \SHOES{} prior \cite{Poulin:2018cxd,Smith:2019ihp,Murgia:2020ryi}. Furthermore, to illustrate the role of {\em Planck} polarization data at the spectrum level, we compare in Fig. \ref{fig: residuals_full} the TT, TE and EE power spectra between the EDE and \LCDM{} best-fit models obtained when analyzing \ACT{} and SPT-3G data with and without {\em Planck} polarization data.

\begin{table*}[h!]
\def\arraystretch{1.2}
 \scalebox{1.0}{
 \begin{tabular}{|l|c|c|c|} 
    \hline Parameter & \textit{Planck} TT650+$\tau$ & \textit{Planck} TT650+$\tau$ & \textit{Planck} TT650+$\tau$\\
    & +ACT DR4 & +SPT-3G & +ACT DR4+SPT-3G\\
    \hline \hline
    $f_{\rm EDE}(z_c)$  &  $0.121(0.113)_{-0.047}^{+0.029}$  & $<0.203 (0.148)$&  $0.102(0.099)_{-0.057}^{+0.034}$ \\
    $\log_{10}(z_c)$&  $3.208(3.221)_{-0.095}^{+0.11}$&$3.46(3.56)_{-0.2}^{+0.19}$ & $3.256(3.295)_{-0.1}^{+0.15}$ \\
    $\theta_i$ &unconstrained (0.561) & unconstrained (2.623)& unconstrained (0.474)  \\
    \hline
    $H_0$ [km/s/Mpc] & $73.22(73.36)_{-3}^{+2.3}$  &$72.3(73.89)_{-3.6}^{+1.9}$ & $72.94(73.26)_{-2.8}^{+2.7}$  \\
    $100~\omega_b$&  $2.154(2.145)_{-0.047}^{+0.04}$& $2.272(2.290)_{-0.044}^{+0.039}$  & $2.219(2.216)_{-0.038}^{+0.03}$   \\
    $\omega_{\rm cdm}$& $0.1308(0.1303)_{-0.0092}^{+0.0053}$ & $0.1233(0.1304)_{-0.01}^{+0.005}$& $0.1279(0.1283)_{-0.0086}^{+0.0049}$  \\
    $10^9A_s$ & $2.121(2.118)_{-0.062}^{+0.048}$ &$2.091(2.105)_{-0.051}^{+0.043}$ & $2.119(2.121)_{-0.042}^{+0.041}$ \\
    $n_s$&  $0.9785(0.9781)_{-0.017}^{+0.012}$ &$0.9878(0.9964)_{-0.018}^{+0.011}$ &$0.988(0.9883)_{-0.017}^{+0.012}$   \\
    $\tau_{\rm reio}$&  $0.0547(0.0547)_{-0.0076}^{+0.0075}$ &$0.0541(0.0532)_{-0.0074}^{+0.0075}$ &$0.0546(0.0546)_{-0.0071}^{+0.0073}$   \\
    \hline
    %$M_b$ &  & & & \\
    $S_8$ & $0.809(0.806)_{-0.048}^{+0.044}$  &  $0.782(0.797)\pm0.043$& $0.803(0.804)_{-0.032}^{+0.034}$ \\
    $\Omega_m$ &  $0.286(0.283)_{-0.019}^{+0.018}$ &$0.281(0.282)_{-0.02}^{+0.018}$ & $0.284 (0.282)_{-0.017}^{+0.015}$\\
    Age [Gyrs] &  $13.06(13.05)_{-0.24}^{+0.42}$ & $13.29(12.97)_{-0.19}^{+0.49}$& $13.14 (13.09)_{-0.33}^{+0.39}$  \\
    \hline
    $\Delta\chi^2_{\rm min}({\rm EDE}-\Lambda{\rm CDM})$ & -16.9& -5.3 & -9.9 \\
    %Preference over $\Lambda$CDM & 99.93\% (3.4$\sigma$) &84.9\% (1.4$\sigma$) & 98.1\% (2.3$\sigma$) \\
    Preference over $\Lambda$CDM & 3.4$\sigma$ & 1.4$\sigma$ & 2.3$\sigma$ \\
    \hline
\end{tabular} }
\caption{The mean (best-fit) $\pm 1\sigma$ errors of the cosmological parameters reconstructed from analyses of various data sets (see column title) in the EDE model. For each data set, we also report the best-fit $\chi^2$ and the $\Delta\chi^2\equiv\chi^2(\Lambda{\rm CDM})-\chi^2({\rm EDE})$.}
\label{tab:TTonly}
\end{table*}

\FloatBarrier
\newpage

\section{Supplementary material on tests for systematic errors within {\em Planck} polarization data}
\label{app:syst}
\begin{figure*}
    \centering
    \includegraphics[width=0.6\textwidth]{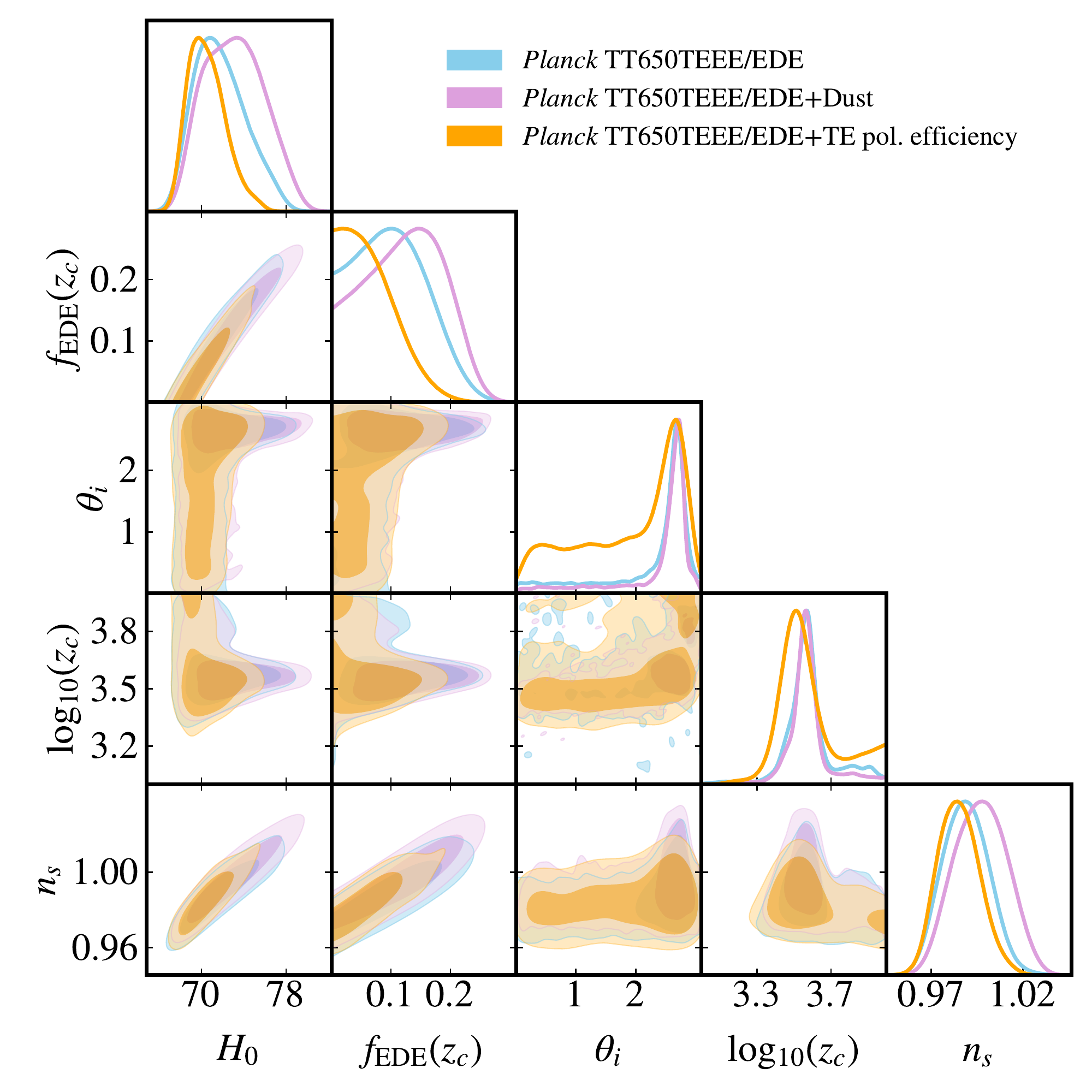}
    \caption{1D and 2D posterior distributions (68\% and 95\% CL) of $\{H_0, f_{\rm EDE}(z_c), \theta_i, \log_{10}(z_c), n_s\}$ reconstructed from {\em Planck} TT650TEEE when considering two potential type of systematic errors (\textit{Planck} TE polarization efficiency and dust contamination) compared to the fiducial run. }
    \label{fig:syst}
\end{figure*}

We present in this appendix the results of two tests for systematic errors within {\em Planck} TEEE data. First, we test a different approach for the modeling of the {\em Planck} TE polarization efficiency (PE) calibration and, second, we test the impact of galactic dust contamination amplitudes \cite{Planck:2019nip} (see Section \ref{sec:systs} for more details on these two sources of systematics). We show the result of our analyses of {\em Planck} TT650TEEE+low-${\ell}$ TTEE data in Fig.~\ref{fig:syst}, where we plot the reconstructed 1D and 2D posterior distributions for the most relevant parameters for our discussion. As was discussed in the main text, one can see that if the PE is chosen to be different than the baseline, the posterior for $f_{\rm EDE}(z_c)$ becomes compatible with zero, and we derive $f_{\rm EDE}(z_c)<0.151$ (95\%C.L.) with $\Delta\chi^2=-5.1$. On the other hand, placing uniform priors on the dust contamination amplitude does not alter our results, as we reconstruct $f_{\rm EDE}(z_c) = 0.131_{-0.049}^{+0.085}$ with $\Delta\chi^2=-10.2$.

\FloatBarrier
\newpage

\section{Supplementary material on results with a \SHOES{} prior on $H_0$}
\label{app:ext_full}
In this appendix we present results of the analyses that include a late-time prior on $H_0$ as measured by \SHOES{}. We perform one analysis that includes {\it Planck} TT650TEEE, \ACT{}, SPT-3G, BAO and Pantheon data, and another that consider the full {\it Planck} temperature power spectrum, as well as $f\sigma_8$ and {\em Planck} lensing ($\phi\phi$) data. A complete list of constraints is given in Table~\ref{tab:H0} and the $\chi^2$ per experiments are reported in Table.~\ref{tab:chi2_h0}.

\begin{table*}[h!]
\def\arraystretch{1.2}
 \scalebox{1.0}{
 \begin{tabular}{|l|c|c|} 
   \hline Parameter & \textit{Planck} TT650TEEE & \textit{Planck} TTTEEE\\
    &+ACT DR4+SPT-3G &+ACT DR4+SPT-3G+$\phi\phi$\\
    &+BAO+Pantheon+\SHOES{} & +BAO/$f\sigma_8$+Pantheon+\SHOES{}\\
   \hline \hline
    $f_{\rm EDE}(z_c)$  &  $0.143(0.153)_{-0.026}^{+0.023}$&$0.116(0.115)_{-0.022}^{+0.023}$  \\
    $\log_{10}(z_c)$ & $3.523(3.525)_{-0.027}^{+0.032}$ & $3.543(3.510)_{-0.036}^{+0.031}$  \\
    $\theta_i$  & $2.731(2.761)_{-0.061}^{+0.098}$ & $2.75(2.81)_{-0.06}^{+0.09}$  \\
    \hline
    $H_0$ [km/s/Mpc] &$72.81(73.08)_{-0.98}^{+0.82}$  & $71.68(71.23)_{-0.85}^{+0.77}$  \\
    $100~\omega_b$ &  $2.275(2.268)_{-0.019}^{+0.017}$& $2.264(2.242)_{-0.014}^{+0.014}$ \\
    $\omega_{\rm cdm}$ & $0.1342(0.1355)_{-0.0041}^{+0.0032}$ &$0.131(0.132)_{-0.003}^{+0.003}$  \\
    $10^9A_s$ &$2.137(2.140)_{-0.035}^{+0.038}$  & $2.147(2.125)_{-0.028}^{+0.026}$  \\
    $n_s$ & $0.9956(0.9955)_{-0.006}^{+0.0057}$ &  $0.9887(0.9833)_{-0.0058}^{+0.0052}$ \\
    $\tau_{\rm reio}$ & $0.0509(0.0508)_{-0.0076}^{+0.0082}$  &  $0.0543(0.0477)_{-0.0066}^{+0.0068}$\\
    \hline
    $S_8$  & $0.838(0.841)_{-0.016}^{+0.015}$ & $0.839(0.841)\pm0.011$  \\
    $\Omega_m$ & $0.297(0.297)_{-0.006}^{+0.005}$  & $0.300(0.305)\pm0.005$   \\
    Age [Gyrs]  &  $12.98(12.93)_{-0.13}^{+0.16}$  &  $13.14(13.17)_{-0.13}^{+0.12}$\\
    \hline
    $\Delta \chi^2_{\rm min}({\rm EDE}-\Lambda{\rm CDM})$ & -34.6 & -25.2 \\
    %Preference over $\Lambda$CDM & 99.999985\% (5.3$\sigma$) &99.9986\% (4.3$\sigma$) \\
    Preference over $\Lambda$CDM & 5.3$\sigma$ & 4.3$\sigma$ \\
    \hline
\end{tabular} }
\caption{The mean (best-fit) $\pm 1\sigma$ errors of the cosmological parameters reconstructed from analyses of various data sets (see column title) in the EDE model when including a late-time prior on the Hubble parameter following the latest value reported by \SHOES{}. For each data set, we also report the best-fit $\chi^2$ and the $\Delta\chi^2\equiv\chi^2({\rm EDE})-\chi^2(\Lambda{\rm CDM})$.}
\label{tab:H0}
\end{table*}

\begin{table*}[h!]
\def\arraystretch{1.2}
\scalebox{1}{
\begin{tabular}{|l|c|c|c|c|}
    \hline
    & \multicolumn{2}{|c|}{$\Lambda$CDM}& \multicolumn{2}{|c|}{EDE} \\
    \hline
    {\emph{Planck}}~high$-\ell$ TT650TEEE & 1843.4 &$-$ & 1836.1&$-$\\
    {\emph{Planck}}~high$-\ell$ TTTEEE &$-$ &2351.0 &$-$ &2348.5\\
    {\emph{Planck}}~low$-\ell$ TT & 21.0 & 21.8& 20.6 &21.7\\
    {\emph{Planck}}~low$-\ell$ EE &395.8 & 396.9&395.8 &395.9\\
    {\emph{Planck}}~lensing& $-$& 9.0&$-$ &9.9\\
    ACT DR4& 297.8& 242.5& 288.5&236.8\\
    SPT-3G & 519.1&520.3 &518.4 &520.1\\
    BOSS BAO low$-z$ & 2.4&1.9 & 2.2&1.6\\ 
    BOSS BAO DR12 & 3.7&$-$ &3.5 &$-$\\ 
    BOSS BAO/$f\sigma_8$ DR12 & $-$& 6.0&$-$ &6.6\\
    Pantheon & 1026.7& 1026.7& 1026.7& 1026.7\\
    \SHOES{}  &16.5 & 19.9 &0.002 &3.0\\ 
    \hline
    total $\chi^2_{\rm min}$& 4126.4& 4596.0& 4091.8&4570.8\\
    $\Delta \chi^2_{\rm min}({\rm EDE}-\Lambda{\rm CDM})$%&~~~~~~~~~~~~$-$~~~~~~~~~~~~~&~~~~~~~~~~~~$-$~~~~~~~~~~~~~& -34.6&-25.2\\
    & $-$ & $-$ & -34.6&-25.2\\
    \hline
    %Preference over $\Lambda$CDM& $-$&$-$ &99.999985\% (5.3$\sigma$) & 99.9986\% (4.3$\sigma$) \\
    Preference over $\Lambda$CDM& $-$ & $-$ & 5.3$\sigma$ & 4.3$\sigma$ \\
    \hline
\end{tabular}}
\caption{Best-fit $\chi^2$ per experiment (and total) for \LCDM{} and EDE when fit to different data combinations including a prior on the $H_0$ parameter as measured by \SHOES{}. We also report the $\Delta \chi^2_{\rm min}\equiv\chi^2_{\rm min}({\rm EDE})-\chi^2_{\rm min}(\Lambda{\rm CDM})$ and the corresponding preference over $\Lambda$CDM, computed assuming the $\Delta\chi^2$ follows a $\chi^2$-distribution with three degrees of freedom.  }
\label{tab:chi2_h0}
\end{table*}

\end{document}